\documentclass[10pt,final,journal,letterpaper,oneside,twocolumn]{IEEEtran}

\usepackage{fixltx2e}

\usepackage[cmex10]{amsmath}

\usepackage[all]{xy}

\usepackage{amsfonts}
\usepackage{amssymb}

\usepackage{graphicx}
\usepackage{enumerate}
\usepackage{url}
\usepackage{tikz}
\usepackage{subfig}

\newcommand{\BB}{\mathbb{B}}
\newcommand{\JJ}{\mathcal{J}}
\newcommand{\II}{\mathcal{I}}
\newcommand{\KK}{\mathcal{K}}
\newcommand{\Fou}{\mathcal{F}}
\newcommand{\qed}{\hspace*{\fill}~{\rule{2mm}{2mm}}\par\endtrivlist\unskip}

\newcommand{\Y}{\mathbb{Y}}
\newcommand{\Z}{\mathbb{Z}}

\newcommand{\I}{\mathcal{I}}
\newcommand{\J}{\mathcal{J}}
\newcommand{\A}{\mathcal{A}}
\newcommand{\B}{\mathcal{B}}
\newcommand{\E}{\mathcal{E}}
\newcommand{\LL}{\mathcal{L}}
\newcommand{\RR}{\mathcal{R}}
\newcommand{\XX}{\mathcal{X}}
\newcommand{\YY}{\mathcal{Y}}
\newcommand{\ZZ}{\mathcal{Z}}

\newcommand{\uFT}{{{\mathcal{F}}}}

\newcommand{\supp}{{\rm supp}}

\newcommand{\om}{\omega}

\newcommand{\dl}{\delta}

\newcommand{\Dih}{{\rm Dih}_N}

\newtheorem{lemma}{Lemma}
\newtheorem{theorem}{Theorem}
\newtheorem{corollary}{Corollary}
\newtheorem{proposition}{Proposition}
\newtheorem{remark}{Remark}

\newtheorem{defn}{Definition}

\title{Discrete Sampling and Interpolation:\\ Universal Sampling Sets for Discrete Bandlimited Spaces}

\author{ Brad Osgood \IEEEmembership{Member, IEEE} \quad Aditya Siripuram  \quad William Wu \thanks{Manuscript received August 31, 2011, revised January 20, 2012. A. Siripuram was supported by a Stanford Graduate Fellowship. W. Wu was supported by the Frank and Eva Buck Foundation}\thanks{W. Wu is currently at the Jet Propulsion Laboratory, Pasadena, CA.} \thanks{The authors are listed alphabetically.}\thanks{Copyright (c) 2011 IEEE. Personal use of this material is permitted.  However, permission to use this material for any other purposes must be obtained from the IEEE by sending a request to pubs-permissions@ieee.org.}\\ Information Systems Laboratory\\ Stanford University}

\begin{document}

\maketitle

\begin{abstract}
We study the problem of interpolating all values of a discrete signal $f$ of length $N$ when $d<N$ values are known, especially in the case when the Fourier transform of the signal is zero outside some prescribed index set $\JJ$; these comprise the (generalized) bandlimited spaces $\mathbb{B}^\JJ$. The sampling pattern for $f$ is specified by an index set $\II$, and is said to be a universal sampling set if samples in the locations $\II$ can be used to interpolate signals from $\mathbb{B}^\JJ$ for \emph{any} $\JJ$. When $N$ is a prime power we give several characterizations of universal sampling sets, some structure theorems for such sets, an algorithm for their construction, and a formula that counts them. There are also natural applications to additive uncertainty principles.
\end{abstract}


\section{Introduction} \label{section:introduction}

\IEEEPARstart{I}{n} this paper and in a sequel \cite{OIS} we consider the problem of interpolating all values of a discrete, periodic  signal $f\colon \mathbb{Z}_N \longrightarrow \mathbb{C}$, $N \ge 2$, when $d<N$ values of $f$ are known.  One solution is a discrete form of the classical Nyquist-Shannon theorem, where the spectrum of the signal is assumed to vanish outside a contiguous band of frequencies; see \cite{Will:thesis}, for example.  At the other extreme is the new and important area of compressed sensing, where no assumptions on the spectrum are made. For this, of the many papers we mention only \cite{candes:robust},  \cite{Venk-Bresler} and  \cite{Mishali-Eldar}, since we will refer to this work later. 

Our approach to the problem is in between, though we begin by formulating a very general definition. 
\begin{defn} \label{def:interpolating-system}
Let $\mathbb{Y}$ be a $d$-dimensional subspace of $\mathbb{C}^N$, let $\II \subset [0: N-1]$ be an index set of size $d$, and let $\mathcal{U}_\II=\{u_i \colon i \in \II\}$ be a set of $d$ vectors in $\mathbb{Y}$. We say that $(\II,\mathcal{U}_\II)$ is an \emph{interpolating system} if each $f\in \mathbb{Y}$ can be written as
\begin{equation} \label{eq:interp-formula}
f=\sum_{i\in\II} f(i)u_i.
\end{equation}
We call $\II$ a \emph{sampling set} and $\mathcal{U}_\II$ an \emph{interpolating basis}. When we refer simply to a sampling set we always mean that it is associated with an interpolating basis. If the vectors $u_i$ are orthogonal we say that $(\II,\mathcal{U}_\II)$ is an \emph{orthogonal interpolating system} and that $\mathcal{U}_\II$ is an \emph{orthogonal interpolating basis}. 
\end{defn}

The point of the definition is that the interpolation of all values of $f$ uses the sampled values $f(i)$, $i\in\II$, which might be thought of as measurements of $f$ with respect to the fixed, natural basis of the ambient space $\mathbb{C}^N$, while the basis $\mathcal{U}_\II$ is tailored to $\mathbb{Y}$ and $\II$.\footnote{We could make the definition even more general and allow $\mathbb{Y}$ to be a subspace of any finite-dimensional vector space $\mathbb{X}$, and sample $f\in \mathbb{Y}$ with respect to any fixed basis of $\mathbb{X}$, but the present definition suffices.}  Note that $\II$ need not consist of uniformly spaced indices, so the sampling may be irregular. Indeed, the results described here and in \cite{OIS} were originally motivated by questions from colleagues in medical imaging who had observed that irregular sampling patterns could often give excellent results with less computation. 

For us, to solve the interpolation problem for $\mathbb{Y}$ is to find an interpolating system. It is a linear theory in all aspects. Every subspace has an interpolating system, though it may not be unique, but {not} every subspace has an orthogonal interpolating system. For a given subspace it is also {not} true that any index set is a sampling set for some interpolating basis, so the intervals between samples are not arbitrary. The only subspaces that have \emph{orthonormal} interpolating systems are the coordinate subspaces. All of this is discussed in Section \ref{section:general-properties}.  Orthogonal interpolating systems are the subject of \cite{OIS}, and we find interesting connections with difference sets, perfect graphs, tiling, and we answer affirmatively a discrete version of a conjecture of Fuglede. 

In Section \ref{section:general-properties} we provide some basic results on interpolating systems in general. We quickly move, in Section \ref{section:bandlimited}, to study bandlimited spaces, $\mathbb{B}^\JJ$, defined as signals whose discrete Fourier transforms are supported on $\JJ$. We do not require that $\JJ$ be a set of contiguous indices, so this is more general than the situation in the discrete Nyquist-Shannon theorem (though we continue to use the term ``bandlimited" for short). 

 In Section \ref{section:universal} we begin to concentrate on universal sampling sets, namely index sets $\II$ that are sampling sets for \emph{any} bandlimited space $\mathbb{B}^\JJ$ with $|\JJ|=|\II|$. That is, $\II$ is universal if the sampling pattern specified by $\II$ can be used for interpolation of signals from any $\mathbb{B}^\JJ$. Universal sampling sets were used in \cite{Venk-Bresler} for multicoset sampling and in \cite{Mishali-Eldar} in connection with compressed sensing. Here our central result  gives several necessary and sufficient conditions for an index set to be universal when $N$ is a prime power.   A mathematical consequence of our result is a generalization of Chebotarev's theorem on the invertibility of submatrices of the Fourier matrix.

In Section \ref{section:maximal} we show that a universal sampling set has an interesting structure as a disjoint union of what we call elementary universal sets, and through this analysis we are  able to count the number of  universal sampling sets of a given size. We also introduce maximal (and minimal) universal sampling sets which in turn enter naturally into the uncertainty principles that we discuss in Section \ref{section:uncertainty}.  As an application of uncertainty and universality we prove a ``random'' uncertainty principle, and deduce a generalization of the Cauchy-Davenport theorem from  additive number theory. Our debt to the work in \cite{tao:uncertainty} and \cite{candes:robust}  is clear. Many of our results assume that $N$ is a prime power, and naturally we wonder whether this can be generalized.

The definitions we introduce and the methods we use are based primarily on properties of index sets when the elements are reduced modulo powers of a prime. With a few exceptions (e.g., minimal and cyclotomic polynomials) these can be considered elementary, and it is surprising (to us) how far they lead. The methods here also seem rather different from those of compressed sensing. In compressed sensing, which is nonlinear in theory and practice, the recovery of a signal from samples  does not require knowledge of the frequency spectrum, whereas linear theories like ours cannot do without knowledge of the spectrum.  Nevertheless, with universality the sampling patterns in our approach do not depend on the frequencies, the reconstruction of a signal from its samples is by linear operations, and the samples are ``samples'' in the classical sense instead of random projections of the signal onto a measurement basis as is done in compressed sensing.  Both approaches start with discrete signals, but one needs to sample an analog signal in the first place and this analog sampling generally needs some knowledge of the frequency spectrum.  Works such as \cite{Venk-Bresler} and  \cite{Mishali-Eldar} confront this issue through ``spectrum blind" sampling, and  they end up needing the idea of universality in the process. It is also interesting that the linear theory here can be used to prove a  random uncertainty principle without the necessity of nonlinear techniques, though our result is not as strong as the result in \cite{candes:robust}. We hope to pursue the connections and differences further. We refer to \cite{Will:thesis} and \cite{Aditya:thesis} for additional results, discussion, and examples. See also Appendix \ref{section:additional-references} for references to papers on universality for continuous-time signals.

\section{General Properties, Existence of Interpolating Systems} \label{section:general-properties}

This section is a summary of elementary properties of interpolating systems, including existence theorems in both an algebraic and geometric formulation.   The ideas are simple enough, but they fit together nicely and are an essential foundation for the less simple work to follow.

We fix some notation. Without further comment we will identify a vector in $\mathbb{C}^N$ with its $N$-periodic extension and vice versa, and we typically index vectors from $0$ to $N-1$. (We assume periodicity because the discrete Fourier transform will soon enter the picture.) For $i\in[0 : N-1]$ we let $\dl_i\colon \mathbb{Z}_N\longrightarrow \mathbb{C}$ be the (periodized) discrete $\dl$-function shifted to $i$, so that $\{\dl_0,\dl_1,\dots, \dl_{N-1}\}$ is the natural basis of $\mathbb{C}^N$. The components of a vector in $\mathbb{C}^N$ will always be in terms of the natural basis, but any fixed basis of $\mathbb{C}^N$ would do for the following development. If $\II \subset [0:N-1]$ we let
\[
\mathbb{C}^\II = \rm{span}\{\dl_i\colon i \in \II\}.
\]

Our first goal is to establish
\begin{theorem} \label{theorem:existence}
Any subspace $\mathbb{Y}$ of $\mathbb{C}^N$ has an interpolating system. 
\end{theorem}
We will give two proofs, one geometric and one algebraic, and both are straightforward. 

In the following, $\mathbb{Y}$ is always a subspace of dimension $d$ and $\II$ is always an index set of size $d$. Let $\II'=[0 :N-1]\setminus \II$.  We record several facts.

An interpolating basis for a subspace $\mathbb{Y}$ is trying to be the natural basis in the slots specified by the index set. In fact this is a characterization of interpolating bases. 
\begin{proposition} \label{proposition:characterization}
(i) A basis $\mathcal{U}= \{u_i\colon i\in \II\}$ for $\mathbb{Y}$ is an interpolating basis if and only if 
\[
u_j(i) =\dl_j(i) \quad i,j\in\II.
\]

(ii) Any natural basis vector $\dl_k$ lying in $\mathbb{Y}$ is an element of any interpolating basis of $\mathbb{Y}$. 

(iii) An interpolating basis is determined by its index set, more precisely, if $\{u_i\colon i\in \II\}$ and $\{v_i\colon i\in \II\}$ are interpolating bases for $\mathbb{Y}$ then $u_i=v_i$. for all $i \in \II$
\end{proposition}

Expanding on the first point in Proposition  \ref{proposition:characterization}, the elements of an interpolating basis are perturbations of the natural basis vectors by vectors outside $\mathbb{Y}$:
\begin{proposition} \label{proposition:perturbation}
(i) Any interpolating basis $\{u_i \colon i \in \II\}$ of $\mathbb{Y}$ is of the form
\[
u_i = \dl_i + v_i, 
\]
where $v_i \in \mathbb{C}^{\II'}$. If $v_i \in \mathbb{Y}$ then $v_i=0$. 

(ii) The subspaces of $\mathbb{C}^N$ having an orthogonal interpolating system are of the form $\mathbb{Y}= {\rm span}\{\dl_i+v_i\colon i \in\II\}$ where  the nonzero $v_i$ are orthogonal vectors in $\mathbb{C}^{\II'}$.
\end{proposition}

We omit the proofs of Propositions \ref{proposition:characterization} and \ref{proposition:perturbation}. Part (ii) of Proposition \ref{proposition:perturbation} can be applied in the negative to find examples of subspaces that do not have an orthogonal interpolating basis -- this is a much larger topic -- and it also follows from part (ii) that the only subspaces having an orthonormal interpolating basis are the coordinate subspaces. Both of these points were raised in the introduction.

\begin{IEEEproof}[Geometric Proof of Theorem \ref{theorem:existence}] 
It is easy to see that there is an index set $\JJ$ of size $N-d$ such that $\mathbb{C}^N = \mathbb{Y} \oplus \mathbb{C}^\JJ$. Let $P\colon \mathbb{Y}\oplus \mathbb{C}^\JJ \rightarrow \mathbb{Y}$ be the projection of $\mathbb{C}^N$ onto $\mathbb{Y}$ along $\mathbb{C}^\JJ$. If $f\in \mathbb{Y}$ then, on the one hand,
\[
f = \sum_{i=1}^N f(i) \dl_i.
\]
On the other hand, since $\mathbb{C}^\JJ = {\rm ker} P$ and $Pf = f$ we have
\[
f = Pf = \sum_{i=1}^N f(i)P\dl_i =\sum_{i\not\in\JJ} f(i) P\dl_i.
\]
Thus the $u_i=P\dl_i$ form an interpolating basis of $\mathbb{Y}$ indexed by $\II = [0 : N-1]\setminus \JJ$.
\end{IEEEproof}

We see from this why an interpolating basis need not be unique. The ambiguity in choosing an interpolating basis arises from the ambiguity in choosing a complement; if there is not a unique choice of the complement $\mathbb{C}^\JJ$ of $\mathbb{Y}$, and generally there is not, then there is not a unique interpolating basis for $\mathbb{Y}$. However,  the existence of an interpolating basis  \emph{produces} a complement to $\mathbb{Y}$:
\begin{proposition}
Let $\mathcal{U}=\{u_i\colon i \in \II\}$ be an interpolating basis of $\mathbb{Y}$. Then
$
\mathbb{C}^N = \mathbb{Y} \oplus \mathbb{C}^{\II'}$.
\end{proposition}
\begin{IEEEproof} If we show that $\mathbb{Y}\cap \mathbb{C}^{\II'}=\{{0}\}$ then $\mathcal{U} \cup \{\dl_j\colon j \in \II'\}$ forms a basis for $\mathbb{C}^N$. For this, let $f \in \mathbb{Y}\cap \mathbb{C}^{\II'}$. Then
\begin{equation} \label{eq:f-in-Y}
f = \sum_{i\in \II} f(i)u_i
\end{equation}
because $\mathcal{U}$ is an interpolating basis for $\mathbb{Y}$, and also
\[
f = \sum_{j\in\II'} f(j)\dl_j.
\]
Thus
\[
 \sum_{i\in \II} f(i)u_i = \sum_{j\in\II'} f(j)\dl_j.
\]
Let $k\in \II$ and evaluate both sides at $k$:
\[
\begin{aligned}
 \sum_{i\in \II} f(i)u_i(k) &= \sum_{j\in\II'} f(j)\dl_j(k),\\
f(k) &= 0.
\end{aligned}
\]
By \eqref{eq:f-in-Y}, $f = {0}$ and we are done.
\end{IEEEproof}

The algebraic proof of Theorem \ref{theorem:existence} is in terms of matrices.   
Associate with an index set $\II=\{i_1,i_2,\dots, i_d\}$ the $N \times d$ matrix $E_\II$ whose $d$ columns are the basis vectors $\dl_{i_1}$, $\dl_{i_2}$, \dots, $\dl_{i_d}$. If $R$ is an a $N\times M$ matrix then $E_\II^\textsf{T}R$ is $d \times M$ submatrix of $R$ obtained by choosing the rows indexed by $\II$. In particular, operating by $E_\II^\textsf{T}$ on an $N$-vector $f$ produces the $d$-vector with components $f(i_1)$, $f(i_2)$,\dots, $f(i_d)$. If $R$ is an $M \times N$ matrix then  $RE_\II$ is the $M\times d$ submatrix of $R$ obtained by choosing the columns indexed by $\II$. 

We note three general facts. First,
\(
E_\II^\textsf{T} E_\II = I_d,
\)
where $I_d$ is the $d\times d$ identity matrix. 
Second, if $S$ is a $d\times d$ matrix then 
\(
 E_\II^\textsf{T}(RS)=(E_\II^\textsf{T} R)S\,.
\)
Finally, if $\mathcal{U}=\{u_{i_1},u_{i_2},\dots, u_{i_d}\}$ is a basis for $\mathbb{Y}$ and $U$ is the $N \times d$ matrix whose columns are the $u_i$ then the condition \eqref{eq:interp-formula} that $\mathcal{U}$ be an interpolating basis can be written in matrix form as
\begin{equation} \label{eq:interpolation-formula-1}
f = UE_\II^\textsf{T}f
\end{equation}
for all $f \in \mathbb{Y}$. Here $UE_\II^\textsf{T}$ is an $N \times N$ matrix and we see that 
$\mathcal{U}$ is an interpolating basis for $\mathbb{Y}$ with sampling set $\II$ if and only if $\mathbb{Y} = {\rm ker}(I_N-UE_\II^\textsf{T})$.

Now we have

\begin{IEEEproof}[Algebraic Proof of Theorem \ref{theorem:existence}] 
Take any basis $\mathcal{V}=\{v_1,v_2,\dots,v_d\}$ of $\mathbb{Y}$ and let $R$ be the $N \times d$ matrix whose columns are the basis vectors $v_k$; thus $R_{jk}=v_k(j)$. Since $R$ has rank $d$ it has a $d\times d$ invertible submatrix, and possibly many such submatrices. Let $\II$ be the index set corresponding to the $d$ rows chosen from $R$ to form the invertible submatrix $E_\II^\textsf{T} R$. The columns of the $N\times d$ matrix $R(E_\II^\textsf{T} R)^{-1}$ are again a basis of $\mathbb{Y}$. We write them as $u_{i_1}$, $u_{i_2}$, \dots, $u_{i_d}$, indexed by $\II$. Since
\[
E_\II^\textsf{T}(R(E_\II^\textsf{T} R)^{-1}) = (E_\II^\textsf{T} R)(E_\II^\textsf{T} R)^{-1} = I_d\,,
\]
 the $u_{i_j}$ are as in Proposition \ref{proposition:characterization}, and hence comprise an interpolating basis of $\mathbb{Y}$. 
\end{IEEEproof}

This proof shows how to produce an interpolating basis provided one can find a $d\times d$ invertible submatrix $E_\II^\textsf{T} R$, indexed by $\II$. The more such submatrices the more interpolating bases for $\mathbb{Y}$. On the opposite side, in general not every index set $\II$ is sampling set for an interpolating basis since, in general,  not every choice of a $d \times d$ submatrix is invertible. 

A slightly different way of arranging the algebraic proof also gives an interpolation formula, making \eqref{eq:interpolation-formula-1} more explicit.  As above, let $\mathcal{V}=\{v_1,v_2,\dots,v_d\}$ be a basis of $\Y$ and let $R$ be the corresponding $N \times d$ matrix. If $f \in \Y$ then
\[
f = \sum_{k=1}^N f(k)\dl_k \quad \text{and also}\quad f = \sum_{k=1}^d \alpha_k v_k,
\]
for some constants $\alpha_k$. We want to solve for the $\alpha_k$ in terms of $d$ of the values $f(k)$. 
Write the second equation for $f$ as
\[
f = R{\alpha}, \quad \alpha=(\alpha_1,\alpha_2, \dots, \alpha_d)^\textsf{T}.
\]
Now $R$ has an invertible $d \times d$ submatrix, say $E_\I^\textsf{T} R$ for an index set $\I$, and
so
\[
E_\I^\textsf{T} f = E_\I^\textsf{T}(R {\alpha}) = (E_\I^\textsf{T} R){\alpha}.
\]
We can then solve for ${\alpha}$ via
\[
{\alpha} = (E_\I^\textsf{T} R)^{-1}(E_\I^\textsf{T} f),
\]
resulting in
\begin{equation} \label{eq:interpolation-formula-2}
f = R(E_\I^\textsf{T} R)^{-1}(E_\I^\textsf{T}f).
\end{equation}
This equation writes $f$ in terms of the components $f(i)$, $i \in \I$.

Carrying the algebraic line of reasoning a little further, we also see how two interpolating bases for $\mathbb{Y}$ are related to each other.
\begin{theorem} \label{theorem:matrix-formulation}
Fix an interpolating basis of $\Y$, indexed by $\J$, and let $R$ be the corresponding $N \times d$ matrix. If $S$ is the matrix of another interpolating basis of $\Y$, indexed by $\I$, then $E_\I^\textsf{T} R$ is invertible and
\[
S=R(E_\I^\textsf{T} R)^{-1}\,.
\]
\end{theorem}

\begin{IEEEproof}
Let $\{v_i\colon i\in\I\}$ be the interpolating basis of $\Y$ that are the columns of $S$ and let $\{u_j\colon j\in\J\}$ be the columns of $R$.  Since the $u_j$ are an interpolating basis we can write, for each $i\in \I$,
\[
v_i= \sum_{j\in\J} v_i(j) u_j.
\]
In matrix form this is
\[
S = R(E_\J^\textsf{T} S).
\]
Now multiply on the left by $E_\I^\textsf{T}$, resulting in
\[
E_\I^\textsf{T} S= E_\I^\textsf{T}(R(E_\J^\textsf{T} S)) = (E_\I^\textsf{T} R)(E_\J^\textsf{T} S).
\]
But $E_\I^\textsf{T} S$ is the $d\times d$ identity matrix, so this shows that $E_\I^\textsf{T}R$ is invertible,  that
$(E_\I^\textsf{T} R)^{-1} = E_\J^\textsf{T} S$, and then that
\[
S=R(E_\I^\textsf{T} R)^{-1}.
\]
\end{IEEEproof} 

Finally, we look a little more closely at the interpolating basis provided by 
$R(E_\I^\textsf{T} R)^{-1}$ in relation to the geometric construction.
  From the $d\times  d$ matrix $(E_\I^\textsf{T} R)^{-1}$ form a $d \times N$ matrix by adding $N-d$ columns of zeros in the slots $\I'$. Call this matrix $T$. Then $RT$ is an $N \times N$ matrix and one sees that 
\[
RT\dl_i=
\begin{cases}
u_i,&\quad i \in \I\\
{0},& \quad i \in \I'
\end{cases}
\] 
Thus $RT$ is the projection of $\mathbb{C}^N$ onto $\Y$ along $\mathbb{C}^{\I'}$ and  we are back to the idea of the geometric argument. Observe that whereas the geometric argument started with a complement $\mathbb{C}^{\I'}$ to $\Y$ and produced the interpolating basis via projection, here we started with an interpolating basis for $\Y$ and produced the projection and the complement.

\section{Discrete Bandlimited Spaces} \label{section:bandlimited}

Bandlimited signals are defined by the vanishing of the discrete Fourier transform  outside a set of specified indices. They form a particularly interesting class of subspaces. 

For notation,  let 
\[
\zeta_n= e^{-2\pi i /n},
\]
simplified to just $\zeta$ when $n=N$, and let
$\om\colon\mathbb{Z}_N \longrightarrow \mathbb{C}$ be the discrete complex exponential,
\[
\om(m) = \zeta^m. 
\]
The discrete Fourier transform is then
\[
\uFT f = \sum_{n=0}^{N-1} f(n) \om^{n}.
\]
As usual, we also regard $\uFT$ as an $N \times N$ matrix whose $mn$-entry is $\uFT_{mn}=\om^{n}(m) = \zeta^{mn}$.  We recall that $\uFT^{-1} = (1/N)\uFT^*$ (the adjoint of $\uFT$).

\begin{defn} \label{def:bandlimited}
Let $\JJ\subseteq [0:N-1]$. The $|\JJ|$-dimensional space of bandlimited signals with frequency support $\JJ$ is
\[
\mathbb{B}^\JJ = \uFT^{-1}(\mathbb{C}^\JJ). 
\]
\end{defn}

In words, $f \in \mathbb{B}^\JJ$ if $\uFT f$ has zeros in the slots $\JJ'=[1:N]\setminus \JJ$. There might be more zeros of $\uFT f$ for a given $f$ but there are at least these zeros.  We do not assume that the indices in $\JJ$ are contiguous, so $\uFT f$ is not necessarily supported on a ``band'' of frequencies, but we maintain the use of the term ``bandlimited'' in all cases. 

Since $\uFT^* \dl_n(m) =\zeta^{-mn}$, we get a basis for $\mathbb{B}^\JJ$ by  pulling out of $\uFT^*$ the columns indexed by $\JJ$. Thus we get an interpolating basis with sampling set $\II$ if and only if $E_\II^\textsf{T}\uFT^*E_\JJ$ is invertible, or equivalently if and only if $E_\II^\textsf{T}\uFT E_\JJ$ is invertible. We prefer to use the latter, with $\uFT$ instead of $\uFT^*$.

For the remainder of this paper, interpolating systems for bandlimited spaces will be our main concern. Spaces of bandlimited functions having orthogonal interpolating bases are the subject of \cite{OIS}, but we do have one general observation here: such spaces cannot be too big.
 \begin{proposition} \label{prop:N/2>2}
 If $\mathbb{B}^\JJ$ has an orthogonal interpolating basis then $|\JJ| \le N/2$.
 \end{proposition}
 \begin{IEEEproof}
 Suppose $\mathbb{B}^\JJ$ has an orthogonal interpolating basis indexed by $\II$. Then $|\II|=|\JJ|$. Let $\II'=[0\ : N-1]\setminus \II$. By Proposition \ref{proposition:perturbation} we can write 
 \[
 \mathbb{B}^{\JJ}= {\rm span}\{\dl_i+v_i\colon i \in\II\} ,
 \]
where the $v_i$ are orthogonal vectors in $\mathbb{C}^{\II'}$,  or some possibly $0$. But none of the $v_i$ can be zero, for $\uFT \dl_k = \om^{-k}$ which never vanishes. There are $|\II|$ of the $v$'s, and if $|\JJ|=|\II| > N/2$ then $|\II'| <N/2$ and we would have more than $N/2$ orthogonal vectors in a space of dimension less than $N/2$.
\end{IEEEproof}

\subsection{Necklaces and Bracelets} \label{subsection:necklaces-bracelets}

\begin{figure}
	\centering
		\includegraphics[scale=0.6]{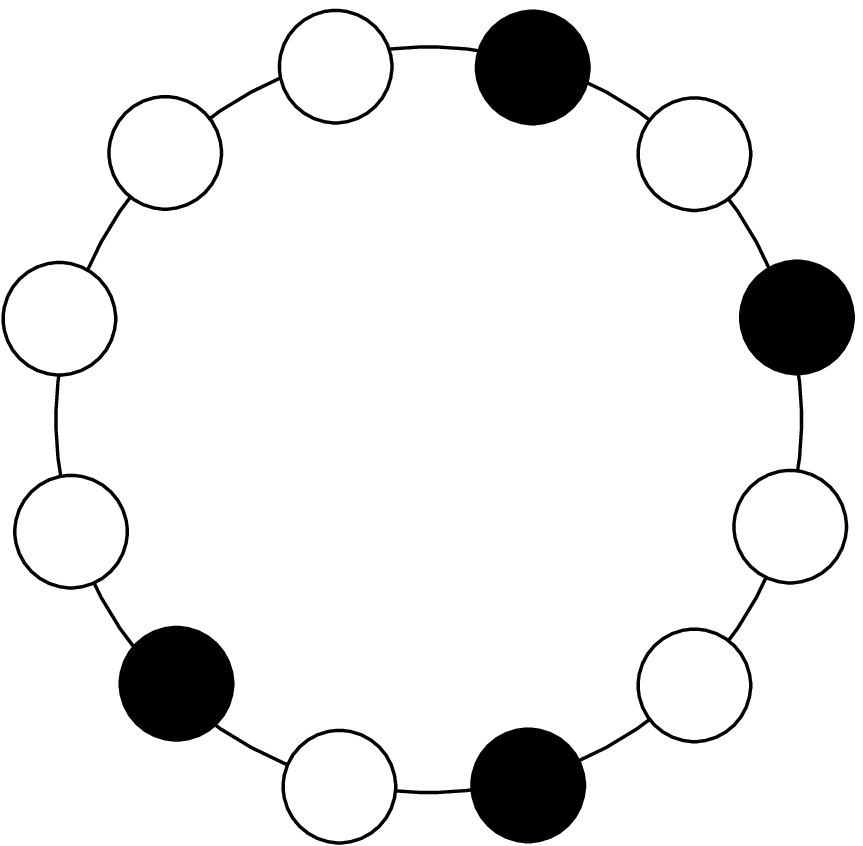} 
		\includegraphics[scale=0.6]{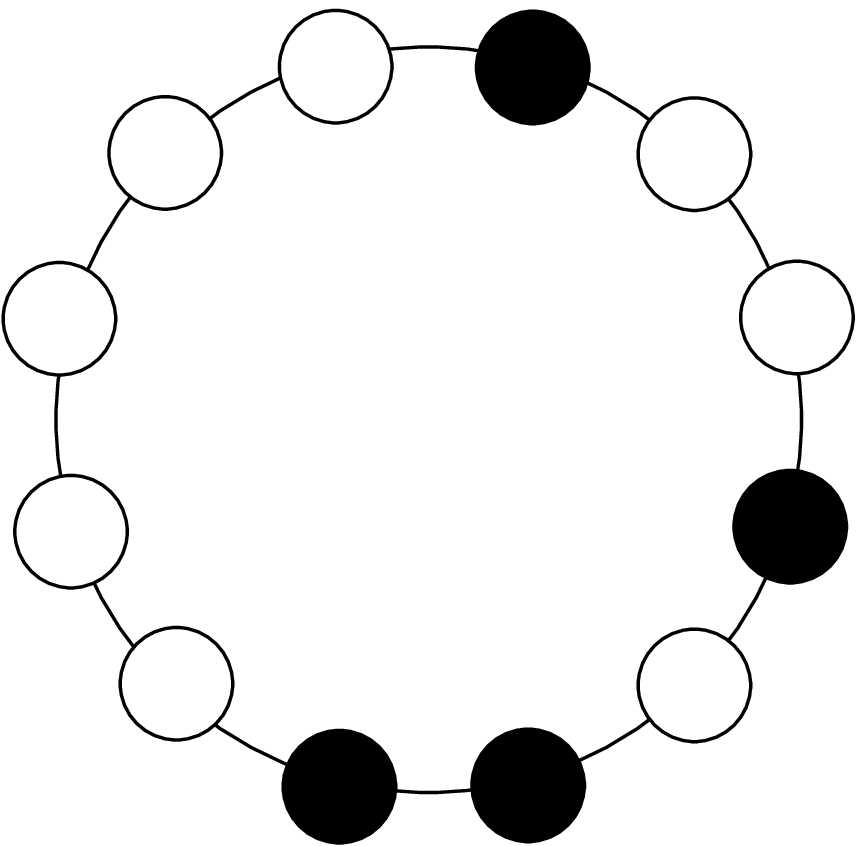}	
        \caption{Two different bracelets with $N=12$ and $|\II|=4$. On top the index set is $\II=\{0,2,5,7\}$, on the bottom the index set is $\II = \{0,3,5,6\}$}
        \label{fig:bracelets}
\end{figure}

 Sampling sets for bandlimited spaces have more algebraic structure than it might appear. Namely, the property of being a sampling set for a particular $\mathbb{B}^\JJ$ is preserved under the action of the dihedral group.  To explain, on $\Z_N$ we denote the operations of {translation} (by $1$) and {reflection} by $\tau$ and $\rho$, respectively:
\[
\begin{aligned}
&\tau \colon \Z_N \longrightarrow \Z_N, \quad \tau(n) = n-1 \mod N,\\
& \rho \colon \Z_N \longrightarrow \Z_N, \quad \rho(n) = -n \mod N .
\end{aligned}
\]
Then 
\[
\tau^N = \rm{id}.\quad \rho^2 = \rm{id} \quad \text{and} \quad \rho\tau\rho = \tau^{-1} \quad \text{or} \quad (\rho\tau)^2 = \rm{id},
\]
so $\tau$ and $\rho$ generate the dihedral group  $\Dih$. Clearly $\Dih$ can act on an index set $\II$ via
\[
\tau\II = \{\tau(i) \colon i \in \II\}, \quad \rho\II = \{\rho(i) \colon i \in \II\}.
\]
We define the \emph{bracelet} of $\I$ to be the orbit of $\I$ under the action of $\Dih$. The \emph{necklace} of $\I$ is  the orbit of $\I$ under the action of the cyclic subgroup $\langle \tau \rangle$ of $\Dih$. Think of $\II\subset [0:N-1]$ as specifying a pattern of $N$ beads on a loop,  with black beads in the locations in $\II$ separated by white beads in the locations in the complement $\II'$, as in Figure \ref{fig:bracelets}.  A necklace is worn around the neck, and if the cyclic group acts then the spacing of the black and white beads is the same however the necklace is rotated. But a bracelet can be worn on either wrist, introducing a reflection, and the symmetry group is $\Dih$. See Appendix \ref{appendix:bracelets} for a formula that counts distinct bracelets, and for references. 

With these definitions we now have
\begin{proposition} \label{prop:bracelet}
If $\II$ is a sampling set for $\mathbb{B}^\J$ then any index set in the bracelet of $\I$ is a sampling set for $\mathbb{B}^\J$.
\end{proposition}

\begin{IEEEproof}
Let $\II = \{m_1, m_2, \ldots, m_d\}$, $\JJ = \{n_1, n_2, \ldots, n_d\}$ and let 
$\KK=\tau\II$. Then the new
submatrix $E_\KK^\textsf{T}\uFT E_\JJ$ is given by

\begin{align*}
\label{eq:brace1}
E_\KK^T\uFT E_\JJ&=
\begin{bmatrix}
\zeta^{ (m_1-1)n_1}  & \zeta^{(m_1-1)n_2}   & \cdots & \zeta^{(m_1-1)n_d}   \\
\zeta^{(m_2-1)n_1} & \zeta^{(m_2-1)n_2} &  \cdots & \zeta^{(m_2-1)n_d}\\
 \vdots & \vdots &  \ddots & \vdots \\
\zeta^{(m_d-1)n_1} & \zeta^{(m_d-1)n_2}
 & \cdots & \zeta^{(m_d-1)n_d}
\end{bmatrix}\\
\vspace{0.1in}
& = \begin{bmatrix}
\zeta^{m_1n_1} & \zeta^{m_1n_2} & \cdots & \zeta^{m_1n_d} \\
\zeta^{m_2n_1} & \zeta^{m_2n_2} &  \cdots & \zeta^{m_2n_d}  \\
 \vdots & \vdots &   \ddots & \vdots \\
\zeta^{m_dn_1}  & \zeta^{m_dn_2}&   \cdots & \zeta^{m_dn_d}
\end{bmatrix}  \times \\
& \hspace{.35in} \begin{bmatrix}
\zeta^{-n_1}& 0 & 0 & \cdots & 0 \\
0 & \zeta^{-n_2} & 0 & \cdots & 0 \\
 \vdots & \vdots & \vdots & \ddots & \vdots \\
 0 & 0 & 0 & \cdots & \zeta^{-n_d}
\end{bmatrix}\\
& = (E_\II^\textsf{T}\uFT E_\JJ) \times\text{an invertible diagonal matrix}.
\end{align*}
Hence $E_\KK^\textsf{T}\uFT E_\JJ$ is invertible whenever $E_\II^\textsf{T}\uFT
E_\JJ$ is, and the same is true for any translation of $\II$.

Next suppose $\KK$ is obtained by reversing $\II$, namely $\KK = \{N-m_1,
N-m_2, \ldots, N-m_d\}$. Then $E_\KK^\textsf{T}\uFT E_\JJ$ is just the
conjugate of $E_\II^\textsf{T}\uFT E_\JJ$, so again, $E_\KK^\textsf{T}\uFT E_\JJ$ is
invertible whenever $E_\II^\textsf{T}\uFT E_\JJ$ is.
\end{IEEEproof}

\section{Universal Sampling Sets} \label{section:universal}

There is a kind of interchange duality for bandlimited spaces between sampling sets and frequency support sets. On the one hand, the sampling problem is to start with $\mathbb{B}^\JJ$ and ask which index sets $\II$ are sampling sets. On the other hand, one could also start with an index set $\II$ and ask which $\mathbb{B}^\JJ$ result from this sampling pattern. These two questions are equivalent.

\begin{proposition} \label{prop:interchange}
$\mathbb{B}^\JJ$ has $\II$ as a sampling set if and only if $\mathbb{B}^\II$ has $\JJ$ as a sampling set.
\end{proposition}

\begin{IEEEproof} The subspace $\mathbb{B}^\JJ$ has $\II$ as a sampling set if and only if $E_\II^\textsf{T}\uFT E_\JJ$ is invertible, and this is true if and only if its transpose $E_\JJ^\textsf{T}\uFT E_\II$ is invertible.
\end{IEEEproof}

Though the sampling problem may seem the more natural one, we will concentrate on the second, equivalent question and ask which frequency patterns, that is which $\mathbb{B}^\JJ$, can arise from a given sampling set $\II$. It may be that the space $\BB^\JJ$ is not known exactly, or that we may have some erroneous estimate $\tilde{\JJ}$ of $\JJ$. The  question is whether we can pick sampling locations $\II$ that are robust for these estimation errors. We will find some interesting phenomena, and the results can easily be translated to apply to the sampling problem.  The extreme case is captured by the following definition.

\begin{defn} An index set $\II \subset [0:N-1]$ is a \emph{universal sampling set}  if $\II$ is a sampling set for each $\BB^{\JJ}$ with $|\JJ| = |\II|$. 
\end{defn}
See also \cite{Venk-Bresler} and \cite{Mishali-Eldar}.

If $\II$ is a universal sampling set, then while an interpolating basis of a space $\mathbb{B}^\JJ$ still depends on $\JJ$, \emph{where} the samples are taken does \emph{not} depend on $\JJ$. In Section \ref{section:maximal} we will show that there are universal sampling sets of any given size; in fact, we will count them.

Very concretely, to ask if $\II$ is a universal sampling set is to ask if there are rows of $\uFT$ indexed by $\II$, $|\II|=d$, such that \emph{any} $d \times d$ submatrix of $\uFT$ formed with these rows is invertible. Phrased this way, standard properties of Vandermonde determinants applied to $\uFT$ allow us to conclude:
\begin{proposition} \label{prop:universal-1}
(i) If $\II$ is a set of $d$ consecutive indices, reduced mod $N$, 
\[
\II=\{i_0,i_0+1,\dots, i_0+(d-1)\} \mod N,
\]
 then $\II$ is a universal sampling set.

(ii) If $\II$ is a set of $d$ indices in arithmetic progression, reduced mod $N$,
\[
\II=\{i_0, i_0+s, i_0+2s,\dots, i_0+(d-1)s\} \mod N, 
\]
where $s$ is coprime to $N$, then $\II$ is a universal sampling set.
\end{proposition} \qed

Much deeper is the following theorem of Chebotarev.

\begin{theorem}
(Chebotarev) If $N$ is prime, then every square submatrix of  $\uFT$ is invertible.
\end{theorem} 

And so, if $N$ is prime then any index set $\II$ is a universal sampling set.  Chebotarev's theorem dates to 1948 (the original paper is in Russian) and there are now several published (and unpublished) proofs, see, e.g., \cite{prasolov:linear-algebra}, \cite{Frenkel:Chebotarev}, but this is by no means a trivial result.

We will generalize Chebotarev's theorem when $N$ is a prime power, and we will offer several characteristic properties of universal sampling sets. We are indebted to the works of Tao \cite{tao:uncertainty} and Delvaux and Van Barel \cite{delvaux:rank-defficient}.

The key is a quantitative, almost statistical comparison of $\II$ to the simplest universal sampling set,
\[
\II^* = [0 : d-1], 
\]
when the  elements of both $\II$ and $\II^*$ are reduced modulo prime powers.  We need several additional definitions to state our main results. 

\subsection{Multisets and the Size of Congruence Classes} \label{subsection:multisets}

We have found it conceptually helpful to use \emph{multisets} in the description of one of the central ideas, and we briefly review this concept. Informally, a multiset is a finite, unordered list $\widetilde{A}$ whose elements are drawn from a finite set $A$, and where, to distinguish a multiset from simply a set,  elements of the list may be repeated. More formally, a multiset  is a pair $(A, \widetilde{\chi}_A)$ where $\widetilde{\chi}_A$ is the multiplicity function (generalizing the characteristic function):
\[
\begin{aligned}
&\widetilde{\chi}_A \colon A \longrightarrow \mathbb{N}, \\
\widetilde{\chi}_A(a) &= \text{the number of times $a\in A$ is listed in $\widetilde{A}$}.
\end{aligned}
\]
Two multisets $\widetilde{A}$ and $\widetilde{B}$ are equal if $\widetilde{\chi}_A = \widetilde{\chi}_B$, so the individual elements are the same and so are their multiplicities. The cardinality of $\widetilde{A}$ is
\[
|\widetilde{A}| = \sum_{a\in A} \widetilde{\chi}_A(a).
\]

 It is common practice to use the standard set notation in writing a multiset. Thus, for example, drawing from $\{a,b,c,d\}$ we write a multiset as $\{ a,a,c,d,d\}$. The tilde notation $\widetilde{A}$ for a multiset drawn from $A$ is helpful in discussing general principles but, like all general notations, it has its limitations in particular cases. It is a notation often used for covering spaces, as we comment on below.

Associated with a multiset $\widetilde{A}$ is another multiset 
\[
\mathcal{M}(\widetilde{A}) = \{\widetilde{\chi}_A(a) \colon a \in A\},
\]
which we call the \emph{multiplicity multiset} of $\widetilde{A}$. Thus $\mathcal{M}(\widetilde{A})$ records as a multiset the counts of the elements of $\widetilde{A}$ and also includes a zero for each element of $A$ that does not appear in $\widetilde{A}$.  One can think of $\mathcal{M}(\widetilde{A})$ as providing some statistics of $\widetilde{A}$, a kind of histogram of $\widetilde{A}$  with bins from $A$, except that the bins are not ordered.

Next, let $p$ be a prime, $k \ge 0$ an integer, and for $x \in \mathbb{N}$ let $[x]_k$  be the 
residue of $x$ reduced mod $p^k$.   
For an index set $\II$ let 
\[
\II/{p^k}=\{[i]_k \colon i \in \II\}
\]
 be the set of  
 residues mod $p^k$ of the elements of $\II$, and let   
 $(\II/p^k)^\sim$ be the corresponding multiset, meaning that each residue is listed according to its multiplicity, i.e, the size of its congruence class. We regard the elements of  $(\II/p^k)^\sim$  
 to be drawn from $[0:p^k-1]$, all possible residues, and we write $\widetilde{\chi}_k \colon [0:p^k-1] \longrightarrow \mathbb{N}$ for the multiplicity function for the multiplicity multiset $\mathcal{M}( (\II/p^k)^\sim)$. To be explicit, for $a \in[0:p^k-1]$
\begin{equation} \label{eq:chi_k}
\begin{aligned}
\widetilde{\chi}_k(a) = &\,\text{the number of elements of $\II$}\\ 
&\,\text{that leave a remainder of $a$ on dividing by $p^k$.}
\end{aligned}
\end{equation}

In particular,  $\widetilde{\chi}_k(a)=0$ means that no element of $\II$ leaves a remainder of $a$ on dividing by $p^k$. In this case we speak of an empty congruence class in $\II/p^k$. For $ a \in [0:p^k-1]$ it will also be helpful to use the notation
\[
\II_{ka} = \{i\in\I \colon i \equiv a \text{ mod  $p^k$}\}    
\]
for the elements of the congruence class of $a$ mod $p^k$.  Then $\widetilde{\chi}_k(a) = |\II_{ka}|$. 

When we need to emphasize the index set, especially in Section \ref{section:maximal}, we will write $\widetilde{\chi}_k(a\,;\II)$. We note the obvious properties:
\begin{itemize}
\item If $\II$ and $\JJ$ are disjoint then $\widetilde{\chi}_k(a\,;\II \cup \JJ) = \widetilde{\chi}_k(a\,;\II) + \widetilde{\chi}_k(a\,;\JJ)$.
\item  $\II \subseteq \JJ \implies \widetilde{\chi}_k(a\,;\II) \le \widetilde{\chi}_k(a\,;\JJ)$.
\end{itemize}

Observe for $k=0$ that $(\II/1)^\sim$ just consists of $|\II|$ zeros and $\widetilde{\chi}_0(0)=|\II|$. More generally,
\begin{equation} \label{eq:cardinality-chi}
| \II | = \sum_{a=0}^{p^k-1} \widetilde{\chi}_k(a).
\end{equation}
 We also note that the multiplicity multiset $\mathcal{M}((\II/p^k)^\sim)$ depends only on the bracelet of $\II$. While the multisets $(\II/p^k)^\sim$ will generally change if $\II$ is shifted or reversed, the counts of the residues on dividing by $p^k$ will be the same:
\begin{equation} \label{eq:multiplicity-bracelet}
\begin{aligned}
\mathcal{M}((\tau\II/p^k)^\sim) & = \mathcal{M}((\II/p^k)^\sim) \quad \text{and}\\
\mathcal{M}((\rho\II/p^k)^\sim) & = \mathcal{M}((\II/p^k)^\sim).
\end{aligned}
\end{equation}

\begin{remark} \label{remark:covering-spaces} Introducing the multiset $(\II/p^k)^\sim$ is reminiscent of introducing covering spaces (for Riemann surfaces) to resolve the problem of multivalued functions. Here we have the remainder map $r\colon\II \longrightarrow \II/p^k$, $r(i)=[i]$, which is generally not injective and so has a multivalued inverse. Think of the residues (with multiplicity) in $(\II/p^k)^\sim$ as tagged by the number they come from, say as a pair $([i],i)$, which serves to distinguish them much as we think of tagging points on different sheets of a covering space of a Riemann surface. Then we have the commutative diagram
\[
\xymatrix{
		& (\II/p^k)^\sim \ar[d]^{\text{pr}}\\
		\II \ar[ur]^{\widetilde{r}} \ar[r]_r &\II/p^k  }
\]		
where $\text{pr}$ is the projection map, $([i],i) \mapsto [i]$ and the  lift $\widetilde{r}(i) =([i],i)$, of $r$  is bijective.  
The value of the multiplicity function $\widetilde{\chi}_k(i)$ is then the number of elements in the preimage $\text{pr}^{-1}([i])$, 
analogous to the number of sheets over $[i]$. It will generally vary with $[i]$.

\end{remark}

Returning to our primary considerations, we write $\widetilde{\chi}_k^*$ to distinguish the special case when $\II=\II^*$.   We will need the following property of $\widetilde{\chi}_k^*$: 
\begin{equation} \label{eq:chi-I*}
|\widetilde{\chi}_k^{*}(a) - \widetilde{\chi}_k^*(b)| \leq 1,
\end{equation}
for all $a,b \in [0:p^k-1]$ and all $k$. In words, when reducing the elements of $\II^*=[0:d-1]$ modulo $p^k$ for any $k$, the conjugacy classes are all of about the same size.  Or, pursuing the analogy above, the preimages $\text{pr}^{-1}([i])$ of the individual residues all have approximately the same number of elements and one might say that $\II^*/p^k$ is \emph{uniformly covered} for each $k$. 

The inequality in  \eqref{eq:chi-I*} is easy to see. For some background calculations we have found it helpful to have a formula  for $\widetilde{\chi}_k^*$ (from which \eqref{eq:chi-I*} also follows).  If $\ell \in \II^*$ with $[\ell]_k= a \in [0:p^k-1]$ then  $\ell = a + \alpha p^k$ for an integer $\alpha \ge 0$, and since $\ell \le d-1$ we must have $0 \le \alpha \le (d-1-a)/p^k$. The number of integers $\alpha$ for which this inequality holds is the number of $\ell$ whose residue is $a$. Thus
\begin{equation} \label{eq:chi^*-explicit}
\widetilde{\chi}_k^*(a) = \left\lfloor \frac{d-1-a}{p^k} +1\right\rfloor.
\end{equation}

\subsection{A Characterization of Universal Sampling Sets} \label{subsection:universal-characterization}

Our main result is:

\begin{theorem} \label{theorem:universal}
Let $\II$ be an index set in $[0:p^M-1]$. The following are equivalent:
\begin{enumerate}
\item[(i)]  $\widetilde{\chi}_k = \widetilde{\chi}_k^*$ 
 for all $0 \le  k \le M$.
\item[(ii)] $|\widetilde{\chi}_k(a)-\widetilde{\chi}_k(b)| \le 1$ for all $a,b \in [0:p^k-1]$ and $0 \le k \le M$.
\item[(iii)] $\II$ is a universal sampling set.
\end{enumerate}
\end{theorem}

According to Proposition \ref{prop:bracelet} and the relations \eqref{eq:multiplicity-bracelet}, any index set in the bracelet of $\II$ is also a universal sampling set. Likewise, any index set in the bracelet of $\II^*$ can serve as a model universal sampling set. Only condition (i) directly compares $\II$ to $\II^*$, and in terms of multisets it could be stated equivalently as
\[
\mathcal{M}(({\II/p^k})^\sim) = \mathcal{M}(({\II^*/p^k})^\sim).
\]
 Condition (i) for $k=0$ guarantees that $\II$ and $\II^*$ have the same size, from \eqref{eq:cardinality-chi}. Computing $ \mathcal{M}(({\II/p^k})^\sim)$ and  $\mathcal{M}(({\II^*/p^k})^\sim)$ for $k \ge M$ is redundant; since all elements in $\II$ and $\II^*$ are in $[0:p^M-1]$, $\mathcal{M}(({\II/p^k})^\sim)$ for $k \ge M$ is just indicative of the cardinality of $\II$ and $\II^*$. Namely, for $k \ge M$,  each of $\mathcal{M}(({\II/p^k})^\sim)$ and $\mathcal{M}(({\II^*/p^k})^\sim)$ contains $|\II|$ ones and $p^k-|\II|$ zeros.  Condition (ii), a property only of $\II$, indirectly compares $\II$ to $\II^*$ via \eqref{eq:chi-I*}. It says that  $\II/p^k$, like $\II^*/p^k$, is uniformly covered for each $k$.

Before we embark on the proof of the theorem, here is an example. Let $N=2^3$, and $\II = \{ 0, 1, 3, 4, 6 \}$. The following are the multisets for $k=1,2,3$:
\[
\begin{aligned}
&({\II/2})^\sim = \{0, 1, 1, 0, 0\},  \quad \mathcal{M}(({\II/2})^\sim)= \{3, 2\}; \nonumber\\
& ({\II/2^2})^\sim= \{0, 1, 3, 0, 2\},  \quad \mathcal{M}(({\II/2^2})^\sim) = \{2, 1, 1, 1\};\nonumber\\
&({\II/2^3})^\sim = \{0, 1, 3, 4, 6\}, \\
& \hspace{.3in} \mathcal{M}(({\II/2^3})^\sim) = \{1, 1, 0, 1, 1, 0, 1, 0\} \nonumber.
\end{aligned}
\]

The computations for $\II^*=\{0,1,2,3,4\}$ yield
\[
\begin{aligned}
&({\II^*/2})^\sim = \{0, 1, 0, 1, 0\}, \quad \mathcal{M}(({\II^*/2})^\sim) = \{3, 2\};\nonumber\\
&({\II^*/2^2})^\sim = \{0, 1, 2, 3, 0\}, \quad  \mathcal{M}(({\II^*/2^2})^\sim) = \{2, 1, 1, 1\}; \nonumber\\
&({\II^*/2^3})^\sim = \{0, 1, 2, 3, 4\}, \\
& \hspace{.3in}  \mathcal{M}(({\II^*/2^3})^\sim) = \{1, 1, 1, 1,1 , 0, 0, 0\}. \nonumber
\end{aligned}
\]
We see that $ \mathcal{M}(({\II/2^k})^\sim) =  \mathcal{M}(({\II^*/2^k})^\sim)$ for $k=1,2,3$, and hence $\II$ is a universal sampling set. So in case the reader has ever wondered, for the $8 \times 8$ Fourier matrix any $5\times 5$ submatrix built from the rows indexed by $\II$, or from the rows of an index set in the bracelet of $\II$, is invertible.

\begin{IEEEproof}[Proof of Theorem \ref{theorem:universal}, (i) $\Longleftrightarrow$ (ii)] Note: This equivalence does not require that $N$ be a prime power.
The implication (i) $\implies$ (ii) is immediate from \eqref{eq:chi-I*}. Assume (ii) holds and 
let 
\[
\chi = \min_a\widetilde{\chi}_k(a).
\]
From (ii) it follows that any $\widetilde{\chi}_k(a)$ is either $\chi$ or $\chi+1$. Suppose $r$ of the $p^k$ numbers
$\widetilde{\chi}_k(a) $ are equal to $\chi + 1$ and the rest
are equal to $\chi$. The cardinality equation, \eqref{eq:cardinality-chi},
\begin{equation}
\label{eq:univ-alt-1}
\sum_{a=0}^{p^k-1}  \widetilde{\chi}_k(a) = |\II|=d,
\end{equation}
 then gives
\[
p^k \chi + r = d, \quad \text{ with } \quad 0\leq r < p^k.
\]
This means that $\chi$ is the quotient on dividing $d$ by $p^k$ and
$r$ is the remainder. In other words, (ii) and
\eqref{eq:univ-alt-1} together uniquely determine the multiset $\mathcal{M}(({\II/p^k})^\sim) = \{
\widetilde{\chi}_k(a)\colon a \in [0,p^k-1] \}$. Since $\II$ and $\II^*$ both satisfy (ii) and
\eqref{eq:univ-alt-1}, we must have $\mathcal{M}(({\II/p^k})^\sim) = \mathcal{M}(({\II^*/p^k})^\sim)$, or $\widetilde{\chi}_k = \widetilde{\chi}_k^*$.
\end{IEEEproof}

We need two lemmas to prove that condition (i) implies that $\II$ is a universal sampling set. The first is a very old theorem on Vandermonde determinants, \cite{Mitchell:determinants},  as updated in \cite{delvaux:rank-defficient}: 

\begin{lemma}[Delvaux and Van Barel]
\label{thm:genvan}
Let 
\begin{equation}
\label{eq:genvan1}
V=
\begin{bmatrix}
 x_1^{m_1} & x_2^{m_1} & x_3^{m_1} & \cdots & x_d^{m_1} \\
 x_1^{m_2} & x_2^{m_2} & x_3^{m_2} & \cdots & x_d^{m_2} \\
 \vdots & \vdots & \vdots & \ddots & \vdots \\
 x_1^{m_d} & x_2^{m_d} & x_3^{m_d} & \cdots & x_d^{m_d}
\end{bmatrix}
\end{equation}
be a $d \times d$ generalized Vandermonde matrix. Then the determinant of $V$ is given by
\begin{equation}
\label{eq:genvan}
\det V = \left(\prod_{i<j}(x_j - x_i)\right)S(x_1, x_2, \ldots, x_d), 
\end{equation}
where $S(x_1, x_2, \ldots, x_d)$ is a symmetric polynomial in $x_1, x_2, \ldots x_d$ with integer coefficients such that 
\[
S(1,1,\ldots,1) =  \frac{\prod_{0\leq i <j \leq d-1}(m_j-m_i)}{\prod_{0\leq i <j \leq d-1}(j-i)}.
\]
\end{lemma}

The polynomial $S$ is called a \emph{Schur polynomial}, see, for example, \cite{stanley:enumerative-combinatorics}. Based on this lemma we deduce a second result that is itself already a sufficient condition for an index set to be a universal sampling set.
\begin{lemma}
\label{lem:univ-suff}
Let $\II = \{m_0, m_1, m_2, \ldots, m_{d-1}\}$. If 
\begin{equation}
\label{eq:magic-frac}
\mu= \frac{\prod_{0\leq i <j \leq d-1}(m_j-m_i)}{\prod_{0\leq i <j \leq d-1}(j-i)}
\end{equation}
is coprime to $p$, then $\II$ is a universal sampling set.
\end{lemma}

Note that without Lemma \ref{thm:genvan}, it would not even be clear that $\mu$ is an integer. An intuitive idea for why this should be so is given below. The proof of Lemma \ref{lem:univ-suff} is along the  lines of the proof of Chebotarev's theorem in \cite{evans:generalized-vandermonde}, and also in \cite{tao:uncertainty}.

\begin{IEEEproof}[Proof of Lemma \ref{lem:univ-suff}]  We make use of Lemma \ref{thm:genvan} in the case when $V = E_{\II}^T \mathcal{F} E_{\JJ} $. Each $x_\ell$ in \eqref{eq:genvan1} is then a power of $\zeta=e^{-2\pi i /N}$, $x_\ell = \zeta^{j_\ell}$, where $\JJ = \{j_1,j_2,\ldots,j_d\}$. 

Suppose $\det V = 0$. From \eqref{eq:genvan}, this means that $S(x_1, x_2, \ldots, x_d) = 0$. Substituting $x_\ell = \zeta^{j_\ell}$ in $S(x_1, x_2, \ldots, x_d) = 0$, we obtain an equation of the form $s(\zeta) = 0$, where $s(x)$ is a polynomial in one variable with integer coefficients. This means that $\zeta$ is a root of $s(x)$ and since $s(x)$ has only integer coefficients, $s(x)$ must contain the minimal polynomial of $\zeta$ over $\mathbb{Z}$ as a factor. 

For $N=p^M$, the minimal polynomial of $\zeta$ over $\mathbb{Z}$ is  $\phi_N(x) = 1 + x^{p^{M-1}} + x^{2p^{M-1}} + x^{3p^{M-1}} + 
\cdots +x^{(p-1)p^{M-1}}$ (the $N$'th cyclotomic polynomial). So we have $\phi_N(x) \mid s(x)$, where
\[
\phi_N(x) = 1 + x^{p^{M-1}} + x^{2p^{M-1}} + x^{3p^{M-1}} + \cdots +x^{(p-1)p^{M-1}} . 
\]
Now $\phi_N$ and $s$ are both polynomials with integer coefficients, hence  $\phi_N(1) \mid s(1)$. However, $\phi_N(1) = p$, and $s(1) = S(1,1,\ldots,,1) = \mu$. Thus 
\[
p \mid  \mu \, \text{ if $\det V = 0$}.
\]
This proves the lemma.
\end{IEEEproof}

Chebotarev's theorem follows from this result. If $N$ is a prime $p$ then $\mu$ is coprime to $p$ because every factor in the numerator and denominator of $\mu$ is an integer strictly between $-p$ and  $p$. 

We can now complete the proof of one direction of the implications in Theorem \ref{theorem:universal}. 
\begin{IEEEproof}[Proof of Theorem \ref{theorem:universal}: (i) $\implies$ (iii)] 
Let $\II = \{m_1,m_2,m_3,\ldots,m_d\}$ and consider the product of differences
\[
A=\prod_{1\leq i < j \leq d}(m_j - m_i). 
\]
 
There are $\widetilde{\chi}_k(\ell)$ elements of $\II$ that leave a remainder of $\ell$ when divided by $p^k$. Moreover, $m_i \equiv m_j \mod p^k$ if and only if  $p^k \mid (m_j - m_i)$. The number of differences that have a factor of $p^k$ (or higher: $p^r$ for $r>k$) is 
\[
\sum_{l=0}^{p^k-1} \binom{\widetilde{\chi}_k(l)}{2},
\]
and hence the number of differences that have a factor of exactly $p^k$ is given by 
\[
\sum_{l=0}^{p^k-1} \binom{\widetilde{\chi}_k(l)}{2} - \sum_{l=0}^{p^{k+1}-1} \binom{\widetilde{\chi}_{k+1}(l)}{2}.
\]
The largest power of $p$ that divides $A$ is then $p$ raised to
\begin{equation} \label{eq:largest-power-of-p}
{\sum_k k\left( \sum_{l=0}^{p^k-1} \binom{\widetilde{\chi}_k(l)}{2} - \sum_{l=0}^{p^{k+1}-1} \binom{\widetilde{\chi}_{k+1}(l)}{2}\right)} .
\end{equation}  
The expression \eqref{eq:largest-power-of-p} depends only on the values of $\widetilde{\chi}_k$, but the hypothesis is that   $\widetilde{\chi}_k = \widetilde{\chi}_k^*$ for $0 \le k \le N$, and therefore the products  $A=\prod(m_j - m_i)$ and $B = \prod(j-i)$ have the same powers of $p$ as factors. Hence  $\mu = A/B$ is coprime to $p$ and from  Lemma \ref{lem:univ-suff} we conclude that $\II$ is a universal sampling set.
\end{IEEEproof}

\begin{remark} 
The argument above  also gives an insight, if not a proof, as to why $\mu=A/B$ in \eqref{eq:magic-frac} is an integer. Suppose  $\mathcal{M}((\II/p^k)^\sim)= \{r_1, r_2, r_3, \ldots, r_d\}$. The power of $p^k$ in $A = \prod (m_i - m_j)$ is given by 
\[
\sum_{i=1}^d {r_i \choose 2} = \frac{1}{2}\left(\sum_{i=1}^d r_i^2 - \sum_{i=1}^d r_i\right).
\]
Now, $\sum_{i=1}^d r_i$ is the cardinality of $\II$ so
\[
\sum_{i=1}^d r_i = d.
\]
Hence for a set $\II$ which has the minimum power of $p^k$ in $A$ it must be that $\mathcal{M}((\II/p^k)^\sim) = \{r_1, r_2,\ldots r_d\}$ is a solution to  
\begin{align*}
&\text{minimize }r_1^2 + r_2^2 + \cdots +r_d^2 \\
&\text{subject to }r_1 + r_2 + \cdots r_d = d.
\end{align*}
On the reals the optimal solution satisfies $r_1 = r_2 = \cdots = r_d$. This suggests that the set $\II$ with the smallest power of $p^k$ in $A$ must have roughly an equal number of elements in each congruence class. $\II^* = \{0,1,2,\ldots, d-1\}$ is one such set. Thus  the power of $p^k$ is smaller in $B = \prod(i-j)$ than in $A = \prod (m_i - m_j)$ for each $p$ and $k$, and, if the reasoning is to trusted, $\mu = A/B$ is an integer.

\end{remark}

To finish the proof of Theorem \ref{theorem:universal} we will derive the following bounds on $\widetilde{\chi}_k$. 

\begin{lemma} \label{lemma:chi-upper-and-lower}
If $\II\subseteq [0:p^M-1]$ is a universal sampling set of size $d$ then 
\begin{equation} \label{eq:chi-upper-and-lower}
\left\lfloor \frac{d}{p^k}\right\rfloor \le \widetilde{\chi}_k(s) \le \left\lceil \frac{d}{p^k}\right\rceil, \quad s\in[0:p^k-1]\,, 0 \le k \le M.
\end{equation}
\end{lemma}

It follows immediately from \eqref{eq:chi-upper-and-lower} that if $\II$ is a universal sampling set then
\[
|\widetilde{\chi}_k(a) - \widetilde{\chi}_k(b)| \le 1, \quad a, b \in [0:p^k-1].
\]
This is condition (ii), and with this result the proof of Theorem \ref{theorem:universal} will be complete.  Incidentally, for the case  $\II=\II^*$, \eqref{eq:chi-upper-and-lower} is a simple consequence of \eqref{eq:chi^*-explicit} and \eqref{eq:chi-I*}.

The argument for Lemma \ref{lemma:chi-upper-and-lower} is through constructing submatrices of the Fourier matrix of known rank to obtain upper and lower bounds for $\widetilde{\chi}_k$. The first step is to build a particular model submatrix, and this requires some  bookkeeping. 

Let $\II \subseteq [0:p^M-1]$,  at this point not assumed to be a universal sampling set. Fix $k\le M$ and $s\in[0:p^k-1]$, and recall that we let
\[
\II_{ks} = \{ i \in \II \colon i \equiv s \text{ mod $p^k$}\}.
\]
The set $\II_{ks}$ has $\widetilde{\chi}_k(s)$ elements. List them, in numerical order, as $i_0, i_1,i_2,\dots, i_c$, where we put $c=\chi_k(s)-1$ to simplify notation. Let $r$ be a positive integer and define the column vector of length $c$ by 
\[
 \mathfrak{z}^r = \begin{bmatrix}
\zeta_{N}^{i_0r} & \zeta_{N}^{i_1r} & \zeta_{N}^{i_2r} & \cdots & \zeta_N^{i_cr}
\end{bmatrix}^\textsf{T}.
\]

Now let $\mathfrak{Z}^r$ be the $c \times p^k$ matrix obtained by repeating $p^k$ copies of the column $\mathfrak{z}^r$:
\[
\mathfrak{Z}^r =\underbrace{\begin{bmatrix} \mathfrak{z}^r  & \mathfrak{z}^r & \mathfrak{z}^r & \cdots \mathfrak{z}^r \end{bmatrix}}_\text{$p^k$ times},
\]
and let $\mathfrak{D}^s$ be the $p^k \times p^k$ diagonal matrix
\[
\mathfrak{D}^s =
\begin{bmatrix}
1 & 0 & 0 & \dots & 0\\
0 & \zeta_{p^k}^{s} & 0  \dots  & 0\\
0 & 0 & \zeta_{p^k}^{2s} & \dots & 0\\
\vdots & \vdots & \vdots & \ddots & \vdots \\
0 & 0 & 0 & \cdots & \zeta_{p^k}^{(p^k-1)s}
\end{bmatrix}  .
\]

Finally,  let $k'=M-k$, and set 
\begin{equation} \label{eq:JJ_k'r}
\JJ_{k'r} = \{0\cdot p^{k'}+r, 1\cdot p^{k'} +r, 2\cdot p^{k'}+r, \dots, (p^k-1)p^{k'}+r\} .
\end{equation}
From the Fourier matrix $\uFT$ we choose $c$ rows indexed by $\II_{ks}$ and $p^k$ columns indexed by $\JJ_{k'r}$. The result of these choices, we claim, results in
\begin{equation} \label{eq:F_rs-1}
E_{\II_{ks}}^\textsf{T} \uFT E_{\JJ_{k'r}} = \mathfrak{Z}^r\mathfrak{D}^s.
\end{equation}

After the preparations, the derivation of \eqref{eq:F_rs-1} is straightforward. The $(a,b)$-entry of $E_{\II_{ks}}^\textsf{T} \uFT E_{\JJ_{k'r}}$ is
\[
\begin{aligned}
\zeta_{N}^{i_a(bp^{k'}+r)}&=\exp\left(-\frac{(2\pi i)i_a(bp^{M-k}+r)}{p^M}\right)\\
&= \exp\left(-\frac{(2\pi i ) i_ab}{p^k}\right)\exp\left(-\frac{(2\pi i )i_ar}{p^M}\right) .
\end{aligned}
\]
But now recall that, by definition, when $i_a\in \II_{ks}$ is divided by $p^k$ it leaves a remainder of $s$, and thus
\[
\begin{aligned}
&\exp\left(-\frac{(2\pi i ) i_ab}{p^k}\right)\exp\left(-\frac{(2\pi i )i_ar}{p^M}\right)\\
 &= \exp\left(-\frac{(2\pi i ) sb}{p^k}\right)\exp\left(-\frac{(2\pi i )i_ar}{p^M}\right) \\
&=\zeta_{p^k}^{sb}\,\zeta_N^{i_a r}.
\end{aligned}
\]

This construction is the basis for the proof of Lemma \ref{lemma:chi-upper-and-lower}, but applied in block form.

\begin{IEEEproof}[Proof of Lemma \ref{lemma:chi-upper-and-lower}]  To deduce the upper bound $\widetilde{\chi}(s) \le \lceil d/p^k\rceil$ we begin by letting
\[
\JJ= \JJ_{k'0}\cup \JJ_{k'1} \cup \JJ_{k'2} \cup \cdots \cup \JJ_{k'd'}, \quad d'=\left\lceil \frac{d}{p^k}\right\rceil -1,
\]
where $\JJ_{k'r}$ is defined as in \eqref{eq:JJ_k'r}. Note that $\JJ$ is a union of $\lceil d/p^k\rceil$ disjoint sets. Each $\JJ_{k'r}'$, $0\le r \le d'=\lceil d/p^k\rceil-1$ indexes the choice of $p^k$ columns from $\uFT$ and applying \eqref{eq:F_rs-1} we have
\[
\begin{aligned}
&E_{\II_{ks}}^\textsf{T}\uFT E_\JJ= \\
&\begin{bmatrix}
E_{\II_{ks}}^\textsf{T}\uFT E_{\JJ_{k'0}} & E_{\II_{ks}}^\textsf{T}\uFT E_{\JJ_{k'1}} & \cdots & E_{\II_{ks}}^\textsf{T}\uFT E_{\JJ_{k'd'}}
\end{bmatrix}\\
&=
\begin{bmatrix}
\mathfrak{Z}^0\mathfrak{D}^s & \mathfrak{Z}^1\mathfrak{D}^s &\cdots & \mathfrak{Z}^{d'}\mathfrak{D}^s
\end{bmatrix}\\
&=
\begin{bmatrix}
\mathfrak{Z}^0 & \mathfrak{Z}^1 &  \cdots & \mathfrak{Z}^{d'}
\end{bmatrix}
\begin{bmatrix}
\mathfrak{D}^s & 0 & \cdots & 0 \\
0 & \mathfrak{D}^s & \cdots &0 \\
\vdots & \vdots & \ddots & \vdots \\
0 & 0 & \cdots & \mathfrak{D}^s 
\end{bmatrix} .
\end{aligned}
\]
The diagonal matrix in this product is invertible, hence
\begin{equation} \label{eq:rank-bound-1}
\begin{aligned}
&\text{Rank of $E_{\II_{ks}}^\textsf{T}\uFT E_\JJ$} = \text{Rank of $\begin{bmatrix}
\mathfrak{Z}^0 & \mathfrak{Z}^1 & \mathfrak{Z}^2 & \cdots & \mathfrak{Z}^{d'}
\end{bmatrix}$}  \\
&\hspace{.5in} \le \text{Number of distinct columns} = \left\lceil\frac{d}{p^k}\right\rceil.
\end{aligned}
\end{equation}
Now, the number of columns of $E_{\II_{ks}}^\textsf{T}\Fou E_{\JJ}$ is equal to 
\begin{align*}
|\JJ| &= |\JJ_{k'0} \cup \JJ_{k'1} \cup \JJ_{k'2} \cup \ldots \cup \JJ_{k'd'}| \\
& = \sum_{r=0}^{ \lceil d/p^k \rceil -1} |\JJ_{k'r}| = p^k\lceil d/p^k \rceil \geq d,
\end{align*}
so there are at least $d$ columns. Hence if $\II$ is a universal sampling set of size $d$ then  $E_{\II}^\textsf{T}\Fou E_{\JJ}$ must be of full row rank. In particular, since $\II_{ks} \subseteq \II$, it must be that 
$E_{\II_{ks}}^\textsf{T}\Fou E_{\JJ}$ is also of full row rank, for each $s$. Next, the number of rows in $E_{\II_{ks}}^\textsf{T}\Fou E_{\JJ}$ is equal to $|\II_{ks}| = \widetilde{\chi}_k(s)$ by definition. From \eqref{eq:rank-bound-1} we know that the rank of $E_{\II}^\textsf{T}\Fou E_{\JJ}$ is at most $\lceil d /p^k \rceil$, and so we have
\[
\text{(Number of rows) $\widetilde{\chi}_k(s)$} \leq \left\lceil \frac{d}{p^k} \right\rceil.
\]

The proof of the lower bound $\widetilde{\chi}_k(s) \ge \lfloor d/p^k \rfloor$ is very similar. This time we construct a set $\JJ$ with $|\JJ| \leq d$, and observe that if $\II$ is a universal sampling set of size $d$, then $E_{\II}^\textsf{T}\Fou E_{\JJ}$ is of full column rank. 

Let
\[
\JJ = \JJ_{k'0} \cup \JJ_{k'1} \cup \JJ_{k'2} \cup \ldots \cup \JJ_{k' d''}, \quad d''=\lfloor d/p^k \rfloor -1.
\] 
Then just as above,
\[
\begin{aligned}
\text{Rank of $E_{\II_{ks}}^\textsf{T}\uFT E_\JJ$} &=  \text{Rank of $\begin{bmatrix}
\mathfrak{Z}^0 & \mathfrak{Z}^1 & \mathfrak{Z}^2 & \cdots & \mathfrak{Z}^{d''}
\end{bmatrix}$}  \\
&\le \text{Number of distinct columns} = \left\lfloor\frac{d}{p^k}\right\rfloor.
\end{aligned}
\]

The number of rows of $E_{\II_{ks}}^\textsf{T}\Fou E_{\JJ}$ is $|\II_{ks}| = \widetilde{\chi}_k(s)$, and so we must have
\begin{equation}
\text{Rank of }E_{\II_{ks}}^\textsf{T}\Fou E_{\JJ} \leq \min \{\lfloor d/p^k \rfloor ,\widetilde{\chi}_k(s)\}. \label{eq:univ-nec-31}
\end{equation} 
Furthermore, 
\[
E_{\II}^\textsf{T}\Fou E_{\JJ} \nonumber\\
= \begin{bmatrix}E_{\II_{k0}}^\textsf{T}\Fou E_{\JJ} \\ E_{\II_{k1}}^\textsf{T}\Fou E_{\JJ} \\
\vdots\\
E_{\II_{k(p^k-1)}}^\textsf{T}\Fou E_{\JJ}\\
 \end{bmatrix} , \nonumber
 \]
whence
\begin{align}
&\text{Row rank of }E_{\II}^\textsf{T}\Fou E_{\JJ} \nonumber \\
&\quad \leq \text{Rank of }E_{\II_{k0}}^\textsf{T}\Fou E_{\JJ} + \text{Rank of }E_{\II_{k1}}^\textsf{T}\Fou E_{\JJ} \nonumber\\
& \hspace{.5in}+ \ldots + \text{Rank of }E_{\II_{k(p^k-1)}}^\textsf{T}\Fou E_{\JJ} \nonumber\\
&\quad \leq \sum_{s=0}^{p^k-1} \min \{\lfloor d/p^k \rfloor ,\widetilde{\chi}_k(s)\} .  \label{eq:univ-nec-32} 
\end{align}

Now, the number of columns indexed by $\JJ$ is $p^k \lfloor d/p^k \rfloor \leq d$. Hence if $\II$ is a universal sampling set of size $d$, we need $E_{\II}^\textsf{T}\Fou E_{\JJ} $ to be of full column rank. From \eqref{eq:univ-nec-32}, this means we must have 
\[
\text{(Number of columns) } p^k \lfloor d/p^k \rfloor \leq \sum_{s=0}^{p^k-1} \min \{\lfloor d/p^k \rfloor ,\widetilde{\chi}_k(s)\}. 
\] 
This inequality will not be satisfied unless $\lfloor d/p^k \rfloor \leq \widetilde{\chi}_k(s)$ for all $s$. This completes the proof. 
\end{IEEEproof}

\begin{remark}
For many values of $d$, it is enough to prove one side of the inequality \eqref{eq:chi-upper-and-lower}. If we know that $\widetilde{\chi}_k(s) \leq \lceil d /p^k \rceil$, then from $\sum_s\widetilde{\chi}_k(s) =d$ and a recurrence relation \eqref{eq:chi-recurrence}, below,  it is possible to prove that $\lfloor d /p^k \rfloor \leq \widetilde{\chi}_k(s)$. Such cases include
\begin{enumerate}
\item $N=p^M$, $d = c_0 p^k + c_1 p^{k-1}$ for $c_0, c_1 \in \{0,1,2,\ldots,p-1\}$.
\item $N=2^M$, $d = c_0 2^k + c_1 2^{k-1} + c_22^{k-2}$ for $c_0, c_1, c_2 \in \{0,1\}$
\item $N=2^M$, $d = 2^k + 2^{k-1} + 2^{k-2} + \ldots + 2^{k-r+1}$ for some $r$,
\end{enumerate}
\end{remark}

\subsection{Digit Reversal and Universal Sampling Sets} \label{subsection:digit-reversal}

There is another interesting characterization of universal sampling sets in terms of \emph{digit reversal}. Expanding in base $p$, any integer $a \in [0:p^m-1]$, $m \ge 1$, can be written uniquely as
\[
a=\alpha_0+\alpha_1p+\alpha_2p^2+\cdots \alpha_{m-1}p^{m-1},
\] 
where the $\alpha$'s are in $[0:p-1]$. We define a permutation $\pi_m\colon [0:p^m-1] \longrightarrow [0:p^m-1]$ by
\[
\begin{aligned}
&\pi_m(\alpha_0+\alpha_1p+\alpha_2p^2+\cdots \alpha_{m-1}p^{m-1}) = \\
&\hspace{.5in} \alpha_{m-1}+\alpha_{m-2}p+\alpha_{m-3}p^2+\cdots +\alpha_0p^{m-1}.
\end{aligned}
\]
The $\alpha$'s are the digits in the base $p$ expansion of $a \in [0:p^m-1]$ and applying $\pi_m$ to $a$ produces the number in $[0:p^m-1]$ with the digits reversed.  For example (an example we will use again  in Section \ref{section:maximal}), take $[0:7]$. Then $\pi_3([0:7])=\{0,4,2,6,1,5,3,7\}$ in that order.  Such digit reversing permutations were used in \cite{delvaux:rank-defficient} to find rank-one submatrices of the Fourier matrix.

The issue for universal sampling sets is how the numbers $\pi_M(\II)$ are dispersed within the interval $[0:p^M-1]$, where, as before, $N=p^M$. To make this precise, take $k\ge 1$ and partition $[0:p^M-1]$ into $p^k$ equal parts:
\[
[0:p^M-1] = \bigcup_{a=0}^{p^k-1}[ap^{k'}:(a+1)p^{k'}-1], \quad k'=M-k.
\] 
For any $\JJ \subseteq [0:p^M-1]$ and $a \in [0:p^k-1]$, let
\[
\phi_k(a\,;\JJ) = |\JJ \cap [ap^{k'},(a+1)p^{k'}-1]|.
\]
We say that $\JJ$ is \emph{uniformly dispersed} in $[0,p^M-1]$ if
\begin{equation} \label{eq;uniformly-dispersed}
|\phi_k(a\,;\JJ)  - \phi_k(b\,;\JJ)| \le 1
\end{equation}
for all $a, b\in[0:p^k-1]$, and $1\le k \le M$.  Thus $\JJ$ is uniformly dispersed if roughly equal numbers
of its elements are in each of the intervals $[ap^{k'}: (a+1)p^{k'}-1]$ for all $1 \le k \le M$, $k'=M-k$. 

We will show 
\begin{equation} \label{eq:phi-chi}
\phi_k(\pi_k(a)\,; \pi_M(\II)) = \widetilde{\chi}_k(a), \quad a \in[0:p^k-1].
\end{equation}
Thus, to the three equivalent conditions in Theorem \ref{theorem:universal} we can add a fourth:
\begin{quote}
\begin{enumerate}
\item[(iv)] $\pi_M(\II)$ is uniformly dispersed. 
\end{enumerate}
\end{quote}

The derivation of \eqref{eq:phi-chi} uses the following lemma.
\begin{lemma}
\label{lemma:univ-alt-2}
If $j \in [0:p^M-1]$ is given by $j = b + ap^{k'}$, $0\leq b \leq
p^{k'}-1 $, then $\pi_M(j) = \pi_k(a) + p^k\pi_{k'}(b)$. 
\end{lemma}

The proof is straightforward, and the argument for \eqref{eq:phi-chi} then goes very quickly. As defined, for any index set $\JJ$, $\phi_k(a\,; \JJ)$ is the number of elements in $\JJ$ that lie in $[ap^{k'} : (a+1)p^{k'}-1]$, and these are precisely the $j \in \JJ$ of the form $ap^{k'}+b$ with $0\le b \le p^{k'}-1$.  Thus for $i \in [0:p^k-1]$,
\[
\begin{aligned}
&{\phi}_k(\pi_k(i)\,;{\pi_M(\II)}) = \\
&\hspace{.25in}\text{the number of }j \in
\pi_M(\II) \\
&\hspace{.35in}\text{ of the form }\pi_k(i) p^{k'} + b, \ 0\leq b \leq
p^{k'}-1 \nonumber \\
&= \text{ number of }j \in
\II \text{ of the form } \\
&\hspace{.35in}p^{k}\pi_{k'}(b) + i, \ 0\leq b \leq
p^{k'}-1 \text{ (from Lemma \ref{lemma:univ-alt-2})} \\
& = \text{ number of }j \in
\II \text{ that leave a remainder of  } \\
&\hspace{.35in} i \text{ on dividing by }p^k
\nonumber \\
& =  \widetilde{\chi}_k(i). \nonumber
\end{aligned}
\]

\section{Structure and Enumeration of Universal Sampling Sets} \label{section:maximal}

 In this section we analyze in detail the structure of universal sampling sets. Specifically we show that when $N=p^M$ is a prime power  such a set $\II$ is the disjoint union of smaller, \emph{elementary universal sets} that depend on the base $p$ expansion of $|\II|$. The method is algorithmic, allowing us to construct universal sets of a given size, and to find a formula that counts the number of universal sets as a function of $p^M$ and $|\II|$. In particular the formula  answers the question:  How likely is it that a randomly chosen index set is universal? Not very likely, but there are several subtle aspects to the answer. For example, we exhibit plots of the counting function showing some striking phenomena depending on the prime $p$.    Our approach is via \emph{maximal universal sampling sets} which, in turn, enter naturally in studying the relationship between universal sampling sets and uncertainty principles. We take up the latter topic in the next section.

\subsection{A Recurrence Relation and Tree for $\widetilde{\chi}$}

When $N=p^M$ the condition that an index set be a universal sampling set depends on the values of $\widetilde{\chi}_k$ for different $k$. To study this we use a recurrence relation in $k$ for $\widetilde{\chi}_k(a)$. The formula holds even when $N$ is not a prime power.  

\begin{lemma} \label{lemma:recursion}
Let $\II \subseteq [0:N-1]$. Then
 \begin{equation} \label{eq:chi-recurrence}
\widetilde{\chi}_{k-1}(a) = \sum_{j=0}^{p-1}\widetilde{\chi}_k(a + jp^{k-1}),  \quad 
 \end{equation}
{for all $a\in[0:p^{k-1}-1]$.}
 \end{lemma}
\begin{IEEEproof}
An integer $x \in \II$ that leaves a remainder of $a$ when divided by $p^{k-1}$ is of the form $x=\alpha p^{k-1} + a$. Let $\alpha = \beta p + \gamma$ for $\gamma \in [0:p-1]$. Then $x = \beta p^k + \gamma p^{k-1} + a$, that is, $x$ leaves a remainder of either $0\cdot p^{k-1}+a, 1\cdot p^{k-1}+a, 2\cdot p^{k-1}+ a,\dots$  or $(p-1)\cdot p^{k-1}+a$ on dividing by $p^k$. The result follows.  
\end{IEEEproof}

When $N=p^M$ the recurrence formula and the relation it expresses between conjugacy classes has an appealing interpretation in terms of a $p$-ary tree. Several arguments in this section will be based on this configuration. 

\begin{figure*}
\centering
\begin{tikzpicture}[scale=1.0]
\node (0) at (0,8) [circle, minimum size = 1.0cm, fill=gray!10, draw] {$\II_{00}$};
\node (1) at (-4,6) [circle, minimum size = 1.0cm, fill=gray!10,
draw] {$\II_{10}$};
\node (2) at (4,6) [circle, minimum size = 1cm, fill=gray!10, draw]
{$\II_{11}$};
\node (3) at (-6,4) [circle, minimum size = 1cm, fill=gray!10, draw]
{$\II_{20}$};
\node (4) at (-2,4) [circle, minimum size = 1cm, fill=gray!10, draw]
{$\II_{22}$};
\node (5) at (2,4) [circle, minimum size = 1cm, fill=gray!10, draw]
{$\II_{21}$};
\node (6) at (6,4) [circle, minimum size = 1cm, fill=gray!10, draw]
{$\II_{23}$};
\node (7) at (-7,2) [circle, minimum size = 1cm, fill=gray!10, draw]
{$\II_{30}$};
\node (8) at (-5,2) [circle, minimum size = 1cm, fill=gray!10, draw]
{$\II_{34}$};
\node (9) at (-3,2) [circle, minimum size = 1cm, fill=gray!10, draw]
{$\II_{32}$};
\node (10) at (-1,2) [circle, minimum size = 1cm, fill=gray!10, draw]
{$\II_{36}$};
\node (11) at (1,2) [circle, minimum size = 1cm, fill=gray!10, draw]
{$\II_{31}$};
\node (12) at (3,2) [circle, minimum size = 1cm, fill=gray!10, draw]
{$\II_{35}$};
\node (13) at (5,2) [circle, minimum size = 1cm, fill=gray!10, draw]
{$\II_{33}$};
\node (14) at (7,2) [circle, minimum size = 1cm, fill=gray!10, draw]
{$\II_{37}$};
\draw (node cs:name=0) -- (node cs:name =1);
\draw (node cs:name=0) -- (node cs:name =2);
\draw (node cs:name=1) -- (node cs:name =3);
\draw (node cs:name=1) -- (node cs:name =4);
\draw (node cs:name=2) -- (node cs:name =5);
\draw (node cs:name=2) -- (node cs:name =6);
\draw (node cs:name=3) -- (node cs:name =7);
\draw (node cs:name=3) -- (node cs:name =8);
\draw (node cs:name=4) -- (node cs:name =9);
\draw (node cs:name=4) -- (node cs:name =10);
\draw (node cs:name=5) -- (node cs:name =11);
\draw (node cs:name=5) -- (node cs:name =12);
\draw (node cs:name=6) -- (node cs:name =13);
\draw (node cs:name=6) -- (node cs:name =14);
\end{tikzpicture}
\caption{A tree representing the relations between the congruence classes, and the recurrence relation satisfied by
          $\widetilde{\chi}_k(a)$. The value of
          $\widetilde{\chi}_k(a)$ at any node is the sum of
          the values of $\widetilde{\chi}_k(a)$ at its children nodes in level $k+1$.}
          \label{fig:tree1}
\end{figure*}
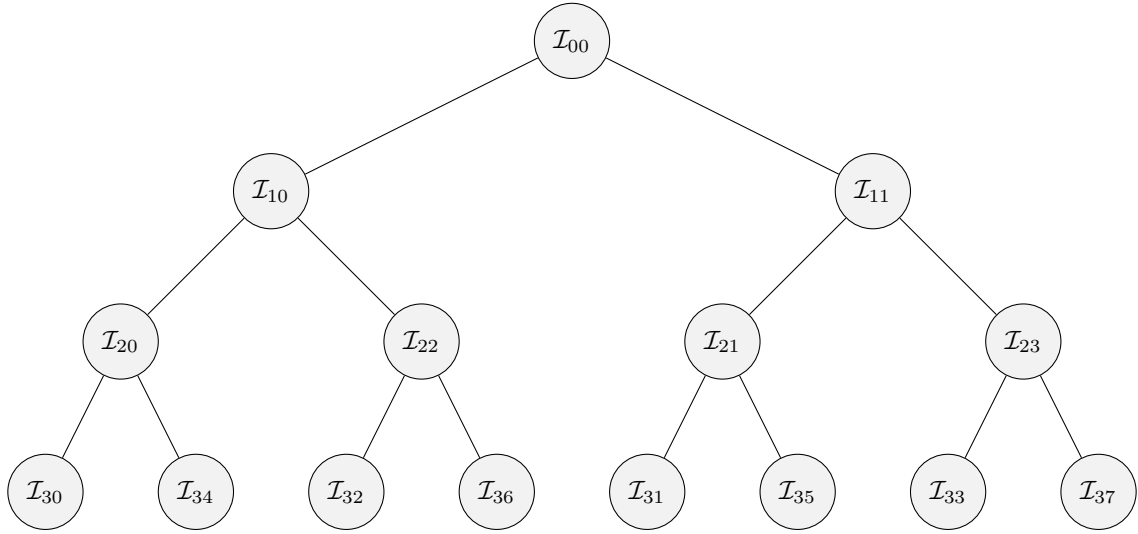

Let $\II\subseteq [0: p^M-1]$. We construct a tree with $M+1$ levels and $p^k$ nodes in level $k$, $0 \le k \le M$. The nodes in level $k$ are identified by a pair $(k,a)$, with $a \in [0:p^{k-1}]$.  Call the
nodes at the level $M$ the leaves. At the node $(k,a)$ we imagine placing the congruence class $\II_{ka} =\{i\in \II \colon i \equiv a \mod p^k\}$. The root is $\II_{00}=\II$ and the nodes at the leaves host the sets $\II_{Ma}$, $a \in [0:p^M-1]$, each of which is either a singleton or empty. We assign a weight of $\widetilde{\chi}_k(a) = |\I_{ka}|$ to the node $(k,a)$. Further, at each level we arrange the nodes according to the digit reversing permutation, i.e., nodes at level $k$ are arranged as $\pi_k([0:p^k-1])$, where $\pi_k$ is the digit reversing permutation from Section \ref{subsection:digit-reversal}. (This is similar to the starting step of the FFT algorithm, where the indices are sorted according to the reversed digits.)  Figure \ref{fig:tree1} shows the case  $N=2^3$, a binary tree with four levels, $k=0, 1, 2, 3$. In  the third level of the tree the nodes are ordered $0, 4, 2, 6, 1, 5, 3, 7$, which is $\pi_3([0:7])$. Then:

\begin{enumerate}
\item The set $\II_{ka}$ at level $k$ is the disjoint union of the sets at its children nodes at level $k+1$.
\item  The value of $\widetilde{\chi}_k(a)$ at the node $(k,a)$ is 
  the sum of the values of $\widetilde{\chi}_{k+1}$ at its children
  nodes at level $k+1$. In other words, the weight of a parent is the sum of the weights of its children; this is the recurrence relation. Consequently, the value of $\widetilde{\chi}_k$ at any
  node is the sum of the values of $\widetilde{\chi}_M$ at the
  leaves at level $M$ descended from the node. 
  \end{enumerate}
  For example, in Figure \ref{fig:tree1} we have
\[
\begin{aligned}
  \widetilde{\chi}_0(0) &= \sum_{a=0}^7\widetilde{\chi}_3(a), \\
 \widetilde{\chi}_1(0) &=\sum_{a=0}^3\widetilde{\chi}_3(2a),\\
 \widetilde{\chi}_1(1) &= \sum_{a=0}^3\widetilde{\chi}_3(2a+1), 
 \end{aligned}
 \]
 and so on.  

In fact, a more general conclusion is the following: Fix a level $k$. Then the value of $\widetilde{\chi}_r$ at any
  node $(r,a)$, for $r \leq k$ is the sum of the values of $\widetilde{\chi}_k$ at the level-$k$ nodes descending from the tree node $(r,a)$.

When the root is $[0:p^M-1]$, the extreme case, the leaves are all singletons and the nodes at level $k$ are each of weight $p^{M-k}$.

\subsection{Elementary and Maximal Sets} \label{subsection:elementary-maximal}

To study the structure of universal sampling sets we need a series of definitions. When $N$ is a prime power the building blocks are the elementary sets: 
\begin{defn} \label{definition:elementary}
A set $\E \subseteq [0:p^M-1]$ is a $k$-\emph{elementary set} if
\[
\widetilde{\chi}_k(a) = 1, \quad \text{~for all $a\in [0:p^k-1]$}.
\]
\end{defn}
Note that $|\E|=p^k$. 

As a first application of the  formula \eqref{eq:chi-recurrence} we can add the adjective ``universal'' to the description of elementary sets.
\begin{lemma} \label{lemma:elementary-universal}
A $k$-elementary set $\E$ is a universal sampling set.
\end{lemma}

\begin{IEEEproof}
From $\widetilde{\chi}_k(a) = 1$ and  \eqref{eq:chi-recurrence} it follows that $\E$ has an equal number of elements in each
congruence class modulo $p^s$, $s \leq k$. More precisely,
\begin{equation} \label{eq:chi-constant}
\widetilde{\chi}_s(a) = p^{k-s},
\end{equation}
for all $s \le k$. Also from \eqref{eq:chi-recurrence}, for $s > k$ all the congruence classes are of size $0$ or $1$, i.e.
\begin{equation} \label{eq:chi-in-{0,1}}
\widetilde{\chi}_s(a) \in\{0,1\}.
\end{equation}
Therefore
\[
|\widetilde{\chi}_s(a)-\widetilde{\chi}_s(b)| \le1,
\]
for all $a,b \in [0: p^k-1]$ and all $s$, and we conclude that $\E$ is a universal sampling set. 
\end{IEEEproof}

Next, a fruitful approach to understanding the structure of universal sampling sets is to ask how well an arbitrary index set is approximated from within by universal sets.

\begin{defn} \label{definition:maximal}
Let $\II \subseteq [0:N-1]$. A \emph{maximal universal sampling set} for $\II$ is a universal sampling set of largest cardinality that is contained in $\II$. 
\end{defn}

Note that the definition does not require $N$ to be a prime power, though this will most often be the case. There is an allied notion of a minimal universal set. We define this in Subsection \ref{subsection:maximal-minimal} below, and show how they are related to maximal sets. Maximal and minimal sets enter naturally and together in connection with uncertainty principles, discussed in Section \ref{section:uncertainty}.

Finding  a maximal universal sampling set for a given $\II$ is a finitary process, so existence is not an issue. However, maximal universal sampling sets need not be unique. For example, take $N=3^2$ and $\mathcal{I}=\{0, 1, 2, 3, 6 \}$. The set $\II$ is not itself a universal sampling set, and both $\{0, 1, 2, 3\}$ and $\{0, 1,2,6 \}$ are maximal universal sampling sets contained in $\mathcal{I}$. 

Despite the lack of uniqueness it will be convenient to have a notation, and we let $\Omega(\II)$ denote a generic maximal universal sampling set in $\II$.  The cardinality $|\Omega(\II)|$ is well-defined; by definition $|\JJ| \le |\Omega(\II)|$ for any universal sampling set $\JJ \subseteq \II$.

Elementary sets and maximal sets are related through an important construction of an elementary set. 

\begin{defn}
Let $\II\subseteq[0:p^M-1]$ and let $\bar{k}$ be the largest integer such that no congruence class in $\II/p^{\bar{k}}$ is empty.  (It might be that $\bar{k}=0$.) Let $\II_{\bar{k}}^\dagger$ denote an elementary set obtained by choosing one element from each congruence class in $\II/p^{\bar{k}}$. 
\end{defn}

By Lemma \ref{lemma:elementary-universal}, $\II_{\bar{k}}^\dagger$ is a universal sampling set, and is of order $p^{\bar{k}}$. 
We now have

\begin{theorem}
\label{thm:maximal-bounds}
Let  $\mathcal{\II} \subseteq [0:p^M-1]$, and $\II_{\bar{k}}^\dagger$ as above. 
Then
\begin{enumerate}
\item[(i)] $p^{\bar{k}}\leq|\Omega(\II) | < p^{\bar{k}+1}$.
\item[(ii)] There exists a maximal universal sampling set contained in $\II$ and containing $\II_{\bar{k}}^\dagger$.
\end{enumerate}
\end{theorem}

\begin{IEEEproof} 
The lower bound in (i) follows from the definition of a maximal set and the comments above,
\[
 p^{\bar{k}} = |\II_{\bar{k}}^\dagger| \le |\Omega(\II)|.
 \]
   To prove the upper bound, suppose  $\JJ \subseteq \II$ has $|\JJ| \geq p^{\bar{k}+1}$. By the
definition of $\bar{k}$ at least one congruence class  in $\JJ/p^{\bar{k}+1}$ is empty, so  $\widetilde{\chi}_{\bar{k}+1}(a\,;\JJ) = 0$ for some $a\in[0:p^{\bar{k}+1}-1]$. From the cardinality equation \eqref{eq:cardinality-chi},
\[
\sum_{\ell=0}^{p^{\bar{k}+1}-1}\widetilde{\chi}_{\bar{k}+1}(\ell \,; \JJ) = |\JJ| \geq p^{\bar{k}+1}.\]
This implies  that at least one congruence class in $\JJ/p^{\bar{k}+1}$ has at least two elements, or $\widetilde{\chi}_{\bar{k}+1}(b\,;\JJ) \geq 2$ for some $b$. We then have
\[
|\widetilde{\chi}_{\bar{k}+1}(b\,;\JJ) - \widetilde{\chi}_{\bar{k}+1}(a\,;\JJ)| = 2 > 1,
\]
and $\JJ$ cannot be a universal sampling set. 

For part {(ii)}, we first show that any maximal universal sampling set $\Omega(\II)$ set must contain at least one element from each congruence class in $\II/p^{\bar{k}}$. By way of contradiction, suppose that $\widetilde{\chi}_{\bar{k}}(a\,;\Omega(\II))=0$ for some $a$. Since $\Omega(\II)$ is universal we must then have $\widetilde{\chi}_{\bar{k}}(b\,;\Omega(\II))\le 1$ for all $b$. By \eqref{eq:cardinality-chi},
\[
|\Omega(\II)|= \sum_{b=0}^{p^{\bar{k}-1}} \widetilde{\chi}_{\bar{k}}(b\,;\Omega(\II)) <p^{\bar{k}},
\]
contradicting the lower bound in (i).

Let $\KK \subseteq \Omega(\II)$ be an elementary set, of size $p^{\bar{k}}$, that contains one element from each congruence class in $\II/p^{\bar{k}}$, guaranteed to exist from what we just showed. Assuming $\KK \ne \II_{\bar{k}}^\dagger$, since otherwise we are done, we will use $\KK$ and $\Omega(\II)$ to construct a (new) maximal universal set that contains $\II_{\bar{k}}^\dagger$. 

Set up a $p$-ary tree, as above, with root $\Omega_{00}=\Omega(\II)$ and  $(\ell,a)$-node the congruence class
\[
\Omega_{\ell a}= \{  i \in \Omega(\II) \colon  i \equiv a \text{ mod $p^\ell$}\},  \quad |\Omega_{\ell a}|=\widetilde{\chi}_\ell(a),  
\]
for $a \in [0:p^\ell-1]$. Recall that $\Omega_{\ell a}$, at level $\ell$, is the disjoint union of the sets at its children nodes at level $\ell+1$. 

Figure \ref{fig:tree2} is an example for $p=3$ and $M\ge 3$, showing only three levels for reasons of space. The shading has to do with the rest of the proof, as we now explain.

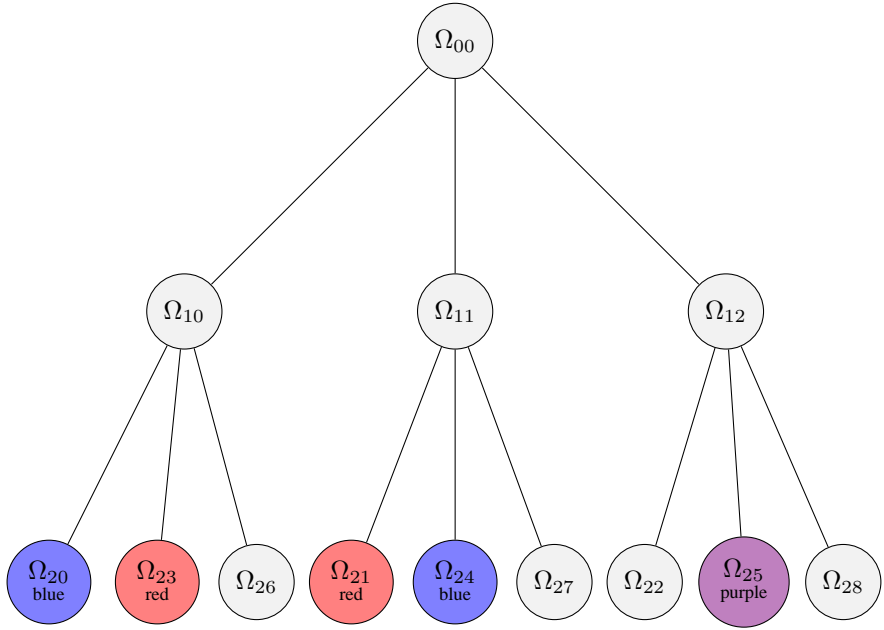
\begin{figure*}
\centering
\begin{tikzpicture}[scale=1.2]
\node (0) at (0,6) [circle, minimum size = 1.0cm, fill=gray!10, draw] {$\Omega_{00}$};
\node (1) at (-3,3) [circle, minimum size = 1.0cm, fill=gray!10,
draw] {$\Omega_{10}$};
\node (2) at (0,3) [circle, minimum size = 1.0cm, fill=gray!10, draw]
{$\Omega_{11}$};
\node (3) at (3,3) [circle, minimum size = 1.0cm, fill=gray!10, draw]
{$\Omega_{12}$};
\node (4) at (-4.5,0) [circle, minimum size = 1cm, fill=blue!50, draw]
{$\underset{{\text{blue}}}{\Omega_{20}}$};
\node (5) at (-3.3,0) [circle, minimum size = 1cm, fill=red!50, draw]
{$\underset{\text{red}}{\Omega_{23}}$};
\node (6) at (-2.2,0) [circle, minimum size = 1cm, fill=gray!10, draw]
{$\Omega_{26}$};
\node (7) at (-1.15,0) [circle, minimum size = 1cm, fill=red!50, draw]
{$\underset{\text{red}}{\Omega_{21}}$};
\node (8) at (0,0) [circle, minimum size = 1cm, fill=blue!50, draw]
{$\underset{\text{blue}}{\Omega_{24}}$};
\node (9) at (1.1,0) [circle, minimum size = 1cm, fill=gray!10, draw]
{$\Omega_{27}$};
\node (10) at (2.1,0) [circle, minimum size = 1cm, fill=gray!10, draw]
{$\Omega_{22}$};
\node (11) at (3.2,0) [circle, minimum size = 1cm, fill=red!50!blue!50, draw]
{$\underset{\text{purple}}{\Omega_{25}}$};
\node (12) at (4.3,0) [circle, minimum size = 1cm, fill=gray!10, draw]
{$\Omega_{28}$};
\draw (node cs:name=0) -- (node cs:name =1);
\draw (node cs:name=0) -- (node cs:name =2);
\draw (node cs:name=0) -- (node cs:name =3);
\draw (node cs:name=1) -- (node cs:name =4);
\draw (node cs:name=1) -- (node cs:name =5);
\draw (node cs:name=1) -- (node cs:name =6);
\draw (node cs:name=2) -- (node cs:name =7);
\draw (node cs:name=2) -- (node cs:name =8);
\draw (node cs:name=2) -- (node cs:name =9);
\draw (node cs:name=3) -- (node cs:name =10);
\draw (node cs:name=3) -- (node cs:name =11);
\draw (node cs:name=3) -- (node cs:name =12);
\end{tikzpicture}
\caption{The congruence class tree for $\Omega(\II)$. The node ${\Omega}_{\ell a}$ is the congruence class of $a$ modulo $p^\ell$ in $\Omega(\II)$, so that $\Omega_{00}=\Omega(\II)$ and $|{\Omega}_{\ell a} | = \widetilde{\chi}_\ell(a)$. }
\label{fig:tree2}
\end{figure*}

Both $\II_{\bar{k}}^\dagger$ and $\KK$ are assembled by choosing single elements from sets at the nodes in the $\bar{k}$-level   (call these the assembly nodes) for a total of $p^{\bar{k}}$ elements for $\I_{\bar{k}}^\dagger$ and $\KK$ each. Observe that the sets at the nodes in the $\bar{k}+1$ level are either empty or singletons. This is so because by definition of $\bar{k}$ there must be some $a\in [0:p^{\bar{k}+1}-1]$ for which $\widetilde{\chi}_{\bar{k}+1}(a) = 0$, and hence by universality $\widetilde{\chi}_{\bar{k}+1}(b) \le 1$ for all $b\in [0:p^{\bar{k}+1}-1]$. And then, according to how the tree is structured, the sets at all nodes farther down in the tree must as well be either empty or singletons.

 Let $\LL \supseteq \II_{\bar{k}}^\dagger$ be the set of elements in $\II$ that leave the same remainders as do the elements in $\II_{\bar{k}}^\dagger$ when divided by $p^{\bar{k}+1}$, more precisely,
\[
\LL= \{j \in \II \colon j \equiv  i \text{ mod $p^{\bar{k}+1}$ for some $i\in \II_{\bar{k}}^\dagger$}\}.
\]
Likewise let  $\LL' \supseteq \KK$ be
\[
\LL'= \{j \in \II \colon j \equiv  i \text{ mod $p^{\bar{k}+1}$ for some $i\in \KK$}\}.
\]
 $\LL$ is the union of the assembly nodes for $\I_{\bar{k}}^\dagger$ and $\LL'$ is the union of the assembly nodes for $\KK$. The collections  may overlap.
 
 We color a node red if it contributes to $\LL$ and blue if it contributes to $\LL'$, and both red and blue (otherwise known as purple) if it contributes to both $\LL$ and $\LL'$.  In the figure we take $\bar{k}=1$, so $\II_{\bar{k}}^\dagger$ and $\KK$ live at the middle level in the tree, as shown.

Focus on each red node in turn. The red node contains an element in $\II^\dagger_{\bar{k}}$, say $i$.
\begin{enumerate}
\item If $\Omega(\II)$ contains an element from this red node, say $j$ (which may or may not be equal to $i$), we replace $j\in \Omega(\II)$ with $i$. This neither changes the size of $\Omega(\II)$ nor the universality.

\item Now suppose $\Omega(\II)$ does not contain an element from this red node. We know that the sibling blue node (i.e. the blue node that shares the parent with this red node) contains an element of $\KK$ (and hence of $\Omega(\II)$), say $j$. Replace $j \in \Omega(\II)$ with $i$. This neither changes the size, nor the universality; we are just exchanging one element from a node with its sibling, so the value of $\widetilde{\chi}$ at the parent node does not change.  
\end{enumerate}

These operations preserve size and universality, and repeating them for each red node ensures that the resultant set contains $\II^\dagger_k$.
 
\end{IEEEproof}

A stronger version of the upper bound in (i)  is the following.
\begin{corollary}
\label{cor:uncer-2bnd}
Let $\II \subseteq[0:p^M-1]$ and let $\underline{k}$ be the smallest integer such that $\widetilde{\chi}_{\underline{k}}(a\,;\II)=0$ for some $a$. Then,
\[
|\Omega(\II)| \leq |\{a: \widetilde{\chi}_{\underline{k}}(a; \II)\neq 0 \}|.
\] 
\end{corollary}
\begin{IEEEproof} From the definition of $\underline{k}$, we have $\widetilde{\chi}_{\underline{k}}(a_0\,; \II) = 0 $ for some $a_0$. Hence by universality, $\Omega(\II)$ must satisfy $|\widetilde{\chi}_{\underline{k}}(b\,; \Omega(\II))| \leq 1$ for all $b$, an observation we used above and will use again. From the cardinality equation \eqref{eq:cardinality-chi}
\[
|\Omega(\II)|  = \sum_b \widetilde{\chi}_{\underline{k}}(b\,; \Omega(\II)) \leq  |\{a: \widetilde{\chi}_{\underline{k}}(a\,; \II)\neq 0 \}|.
\]
\end{IEEEproof}

Ultimately we will show that when $N=p^M$ any maximal universal sampling set, and in particular any universal sampling set, is a disjoint union of elementary sets. In general, however, the union of two disjoint, elementary sets need not be universal. For example, take $N=2^3$, $\E=\{0, 1\}$, $\E'=\{4,5\}$. Then $\E$ and $\E'$ are elementary but their union $\E \cup \E' =\{0,1,4,5\}$ is not universal. What is needed is a kind of independence condition on a collection of elementary sets. The following lemma, whose converse we will also show, makes this latter point precise and introduces the main features of the structure of universal sets. 

\begin{lemma} \label{lem:univ-structure-forward}
Let $N=p^M$. Suppose there exists a finite sequence of nonincreasing integers $k_1 \ge k_2 \ge \cdots \ge 0$ and sets $\E_{r}\subseteq[0:N-1]$, $r =1, 2, \dots$,  such that
\begin{enumerate}[(i)]
\item $\E_{r}$ is $k_r$-elementary.
\item For each $r\ge 1$
\[
\E_{r} \cap \left(\bigcup_{j=1}^{r-1} \LL_j\right) = \emptyset,
\]
where
\end{enumerate}
\[
\mathcal{L}_j = 
\{x  \in [0:N-1] \colon x \equiv e \text{ mod $p^{k_j+1}$ for some $ e \in
\E_{j}$}\} .
\]

Let
\[
\II= \bigcup_r \E_{r}.
\]
Then  $\II$ is a universal sampling set.
\end{lemma}

 Obviously it is condition (ii) that requires further comment. The set $\LL_r$ is defined much as in the proof of Theorem \ref{thm:maximal-bounds}, and we will illustrate the point of (ii) again by means of a tree. Observe first that the $\E_{r}$ are disjoint. This follows from (ii), since $\mathcal{L}_j \supseteq \E_{j}$.

\begin{figure*}
\centering
\begin{tikzpicture}[scale=1.2]
\node (0) at (0,6) [circle, minimum size = 1.0cm, fill=gray!10, draw] {$\mathcal{N}_{00}$};
\node (1) at (-3,3) [circle, minimum size = 1.0cm, fill=gray!10,
draw] {$\mathcal{N}_{10}$};
\node (2) at (0,3) [circle, minimum size = 1.0cm, fill=gray!10, draw]
{$\mathcal{N}_{11}$};
\node (3) at (3,3) [circle, minimum size = 1.0cm, fill=gray!10, draw]
{$\mathcal{N}_{12}$};
\node (4) at (-4.4,0) [circle, minimum size = 1cm, fill=gray!10, draw]
{$\mathcal{N}_{20}$};
\node (5) at (-3.35,0) [circle, minimum size = 1cm, fill=red!50, draw]
{$\underset{\text{red}}{\mathcal{N}_{23}}$};
\node (6) at (-2.3,0) [circle, minimum size = 1cm, fill=gray!10, draw]
{$\mathcal{N}_{26}$};
\node (7) at (-1.1,0) [circle, minimum size = 1cm, fill=red!50, draw]
{$\underset{\text{red}}{\mathcal{N}_{21}}$};
\node (8) at (0,0) [circle, minimum size = 1cm, fill=gray!10, draw]
{$\mathcal{N}_{24}$};
\node (9) at (1,0) [circle, minimum size = 1cm, fill=gray!10, draw]
{$\mathcal{N}_{27}$};
\node (10) at (2.2,0) [circle, minimum size = 1cm, fill=gray!10, draw]
{$\mathcal{N}_{22}$};
\node (11) at (3.3,0) [circle, minimum size = 1cm, fill=red!50, draw]
{$\underset{\text{red}}{\mathcal{N}_{25}}$};
\node (12) at (4.4,0) [circle, minimum size = 1cm, fill=gray!10, draw]
{$\mathcal{N}_{28}$};
\draw (node cs:name=0) -- (node cs:name =1);
\draw (node cs:name=0) -- (node cs:name =2);
\draw (node cs:name=0) -- (node cs:name =3);
\draw (node cs:name=1) -- (node cs:name =4);
\draw (node cs:name=1) -- (node cs:name =5);
\draw (node cs:name=1) -- (node cs:name =6);
\draw (node cs:name=2) -- (node cs:name =7);
\draw (node cs:name=2) -- (node cs:name =8);
\draw (node cs:name=2) -- (node cs:name =9);
\draw (node cs:name=3) -- (node cs:name =10);
\draw (node cs:name=3) -- (node cs:name =11);
\draw (node cs:name=3) -- (node cs:name =12);
\end{tikzpicture}
\caption{Similar to Figure \ref{fig:tree2} but with root $\mathcal{N}_{00}=[0:N-1]$, 
 this tree shows the relationship between $\mathcal{N}_{ka}$, for
  $p=3$, $M\geq 3$.}
  \label{fig:count-tree}
\end{figure*}
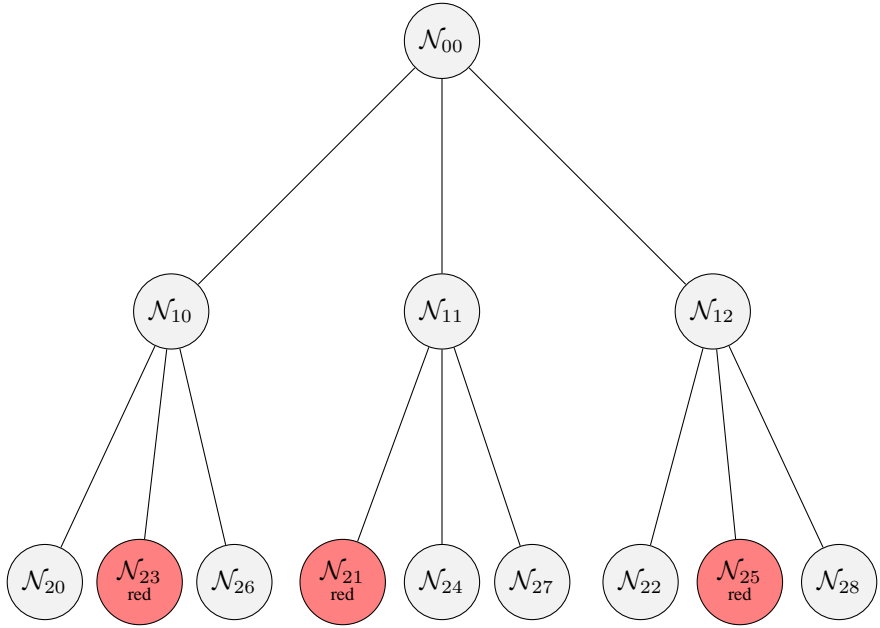

We build a congruence tree with root the full interval $[0:N-1]$. Write this as $\mathcal{N}_{00}$ and write $\mathcal{N}_{ka}$ for the congruence class of $a$ modulo $p^k$ in $[0:N-1]$, so that $|\mathcal{N}_{ka} | = \widetilde{\chi}_k(a\,;[0:N-1])$.  All the nodes represent non-singletons, except the bottom-most level, $M$. As before, Figure \ref{fig:count-tree} has $p=3$, $M\ge 3$ and shows the tree only up to the third level.

  Suppose $k_1 =1$, so $\E_{1}$, as an elementary set, contains one element from each node at the middle  level in the figure. In turn, suppose $\E_1$ comes from picking one element from each of the red nodes. The set $\mathcal{L}_1$ is the union of the red nodes. Now, the set $\E_2$ comes from choosing one element from each node at the $k_2$-level, and the sequence $k_r$ is nonincreasing so $\E_2$ is drawn from nodes in a level at or higher up in the tree than $\E_1$  (in this example $k_2$ is either $1$ or $0$). Condition (ii) requires that $\E_2$ be disjoint from the red nodes, not just from $\E_1$ which is a (small) subset of the red nodes. 
  
  In the general case, think of $k_1$ as large (eventually it will be chosen as in Theorem \ref{thm:maximal-bounds}), so $\E_1$ comes from a level far down the tree from the root, and then $\E_2, \E_3, \dots$ are, at least, no further down since $k_1 \ge k_2 \ge \cdots$. Condition (ii) requires that $\E_r$ be assembled from nodes that were not used in assembling any of the $\E_s$ for $s<r$. It is this property that we exploit to show that $\bigcup_r\E_r$ is universal.

\begin{IEEEproof}[Proof of Lemma \ref{lem:univ-structure-forward}]
Fix $r$ and $s$ with $k_r< s$, and note that 
\begin{equation} \label{eq:chi-in-{0,1}-again}
\widetilde{\chi}_s(a\,;\E_{r}) \in\{0,1\}, \quad  a\in [0:p^{s}-1],
\end{equation}
 from \eqref{eq:chi-in-{0,1}}. Now suppose 
$\widetilde{\chi}_s(a\,;\E_{r}) =1$, so one element in $\E_{r}$ leaves a remainder of $a$ on dividing by $p^s$. Then 
\begin{equation} \label{eq:chi=0-t>r}
\widetilde{\chi}_s(a\,;\E_{t})=0 \quad \text{for all $t>r$},
\end{equation}
 i.e., none of the $\E_{t}$ for $t>r$ will have an element from the congruence class of $a$  modulo $p^s$. This follows (just as described for the tree) from $\E_{t} \cap \LL_r=\emptyset$, and also
\begin{align}
\mathcal{L}_r &= \{x  \in [0:N-1] \colon x \equiv e \text{ mod $p^{k_r+1}$ for some $e \in
\E_{r}$}\}    \nonumber \\
&\supseteq \{x \in [0:N-1]: x \equiv e \text{ mod $p^s$ for some }e \in
\E_{r} \} \nonumber \label{eq:univ-structure-1}.
\end{align}

From \eqref{eq:chi-in-{0,1}-again} and \eqref{eq:chi=0-t>r} we conclude that
\begin{equation} \label{eq:sum-chi-in-{0,1}}
\sum_{r} \widetilde{\chi}_s(a\,;\E_r) \in\{0,1\}
\end{equation}
for all $a$, where the sum is over all $r$ with $k_r <s$. 

With this we can show that $\II = \bigcup_r \E_r$ is universal. For any $s$, and for any $a,b \in [0:p^s-1]$,
\begin{align}
&\widetilde{\chi}_s(a\,;\II) - \widetilde{\chi}_s(b\,;\II)= \sum_r \widetilde{\chi}_s(a\,;\E_r)  - \sum_r \widetilde{\chi}_s(b\,;\E_r) \nonumber \\
& = \left(\sum_{r \,(k_r\geq s)} \widetilde{\chi}_s(a\,;\E_r) - \sum_{r\,(
  k_r\geq s)} \widetilde{\chi}_s(b\,;\E_r)\right) +\\
 & \hspace{.5in} \left(\sum_{r\,(
k_r<s)} \widetilde{\chi}_s(a\,;\E_r) - \sum_{r\,( k_r<s)}
\widetilde{\chi}_s(b\,;\E_r) \right)\nonumber \\
& = \sum_{r\,(k_r<s)} \widetilde{\chi}_s(a\,;\E_r) - \sum_{r\, (k_r<s)}
\widetilde{\chi}_s(b\,;\E_r)\nonumber \\  
&\hspace{.5in}\text{ (the first two sums cancel, by \eqref{eq:chi-constant})}.\nonumber 
\end{align}
From \eqref{eq:sum-chi-in-{0,1}} we have that
$|\widetilde{\chi}_s(a\,;\II) - \widetilde{\chi}_s(b\,;\II)| \leq 1$, 
so $\II$ is universal. 
\end{IEEEproof}

\subsection{An Algorithm to Construct Maximal Universal Sets} \label{subsection:algorithm-maximal}

Consider now the problem of finding
a maximal universal sampling set contained in a  given $\II \subseteq [0:p^M-1]$.  Build the congruence class tree with root $\II$, as in Figure \ref{fig:tree1}, up to level $M$. The leaves having weight $1$ are singletons in $\II$, and $\widetilde{\chi}_k(a\,;\II)$, $a \in [0:p^k-1]$, is the {total weight} at node $(k,a)$.  The problem of constructing $\Omega(\II)$ is to pick a subset of the leaves so that the tree with root $\Omega(\II)$ is well balanced at each level. By `well balanced' we mean that at any given level, all the subtrees have roughly equal weight, corresponding to the condition $|\widetilde{\chi}_k(a\,;\Omega(\II)) -\widetilde{\chi}_k(b\,;\Omega(\II))| \le ~1$. The following algorithm realizes this and provides the value of $|\Omega(\II)|$.  It marries the construction of elementary sets in Theorem \ref{thm:maximal-bounds} with an iterative version of the method used in the proof of Lemma \ref{lem:univ-structure-forward}. 

Let $\II \subseteq [0:p^M-1]$. Initialize with $\II_1 = \II$, and $r=1$. 
\begin{enumerate}
\item Let ${k}_r$ be the largest integer such that no congruence class in $\II_r/p^{{k_r}}$ is empty. 

\item Construct an elementary set $\II_{r}^\dagger \subseteq \II_r$ by choosing one element of $\II_r$ from each congruence class modulo $p^{{k}_r}$. 
(There may not be a unique choice, and this is the reason why there may be many universal sets contained in $\II$.)
\item Define $\LL_r \supseteq \II_r^\dagger$ by \[
\LL_r= \{j \in \II \colon j \equiv  i \text{ mod $p^{{k}_r+1}$ for some $i\in \II_r^\dagger$}\}.
\]

\item Let $\II_{r+1} = \II_r \setminus \LL_r$.  Stop if $\II_{r+1} = \emptyset$. Else increment $r$ to ${r+1}$ and go to (1). 
\end{enumerate}

Note the following:
\begin{enumerate}[(i)]
\item At each step of the algorithm the size of $\II_r$ is reduced by
  $|\LL_r|\geq |\II_r^\dagger|=p^{k_r} \geq 1$:
  \[
  |\II_{r+1}| \leq |\II_r| -
  p^{k_r}.
  \]
   Since $\II= \II_1$ is a
  finite set, the algorithm terminates at some point. 
\item The $k_r$ are nonincreasing: 
\[
 k_1 \geq k_2 \geq k_3 \geq \ldots.
 \]
 \end{enumerate}
 
 We can now state
 \begin{theorem}
\label{theorem:maximal-precise}
With $k_r$, $r \geq 1$, defined as above, we have 
\begin{equation} \label{eq:|Omega(I)|}
|\Omega(\II)| = \sum_r p^{k_r}.
\end{equation}
 One possible maximal universal sampling set is 
\begin{equation} \label{eq:maximal-union}
  \Omega(\II) = \bigcup_r \II_r^\dagger.
\end{equation}
  By construction this is a disjoint union.
\end{theorem}

Here is an example of the algorithm in action. Let $N=2^5$ and 
\[
\II = \{0,1,2,3,4,6,7,8,9,10,12,14, 15\}= \II_1.
\] 
\begin{enumerate}
\item $(r=1)$ Note that $\widetilde{\chi}_3(5\,;\II_1) = 0$, and that no values $\widetilde{\chi}_2(i\,;\II_1)$ are zero. Hence $k_1 = 2$. Form $\II_1^\dagger$ by taking one element from each congruence class in $\II_1$ modulo $2^{k_1} = 4$, e.g. $\II_1^\dagger= \{0,1,2,3\}$.
Then $\LL_1=\{0,1,2,3,8,9,10 \}$ is the set of all elements of $\II_1$ that leave a remainder of $0,1,2$ or $3$ on dividing by $2^{k_1+1} = 8$. Removing such numbers from $\II_1$, we have $\II_2 = \II_1 \setminus \LL_1 = \{4, 6, 7, 12, 14, 15 \}$. 
\item $(r=2)$ Now $\widetilde{\chi}_2(1\,;\II_2) = 0$ while $\widetilde{\chi}_1(0\,;\II_2)$, $\widetilde{\chi}_1(1\,;\II_2) \neq 0$ so $k_2 = 1$. Let $\II_2^\dagger = \{4, 7\}$. Then $\LL_2=\{4,7,12,15\}$ is the set of all elements in $\II_2$ that leave a remainder of $4 \text{ mod }4 = 0 $ or $7 \text{ mod }4 = 3$ on dividing by $2^{k_2+1} = 4$. Removing such numbers from $\II_2$, we have $\II_3 = \II_2 \setminus \LL_2 = \{6, 14 \}$. 
\item $(r=3)$ Now clearly $k_3 = 0$. Let $\II_3^\dagger = \{6\}$. Then $\LL_4 = \{6,14\}$, $\II_4 = \emptyset$ and the algorithm terminates.
\end{enumerate}
According to the theorem, we have $|\Omega(\II)| = 2^{k_1} + 2^{k_2} + 2^{k_3} = 7 $, and an example $\Omega(\II)$ is given by $\II_1^\dagger \cup \II_2^\dagger\cup \II_3^\dagger = \{0,1,2,3,4, 6, 7\}$. 

We have several additional comments. First, we can say more about the formula for $|\Omega(\II)|$. Since the $k_r$'s are nonincreasing,  a typical sequence is, say,
\[
\underbrace{l_1, l_1,\ldots}_{\alpha_1 \text{ times }} \quad \underbrace{l_2, l_2, \ldots}_{ \alpha_2 \text{ times }} \quad \underbrace{l_3, l_3, \ldots}_{ \alpha_3 \text{ times }}, \quad \dots
\]
with $l_1 > l_2 > l_3$. Given this, equation \eqref{eq:|Omega(I)|} appears as 
\begin{equation} \label{eq:Omega(I)-2}
|\Omega(\II)| = \alpha_1p^{l_1} + \alpha_2p^{l_2} + \alpha_3p^{l_3} + \ldots .
\end{equation}
In fact, effectively, Theorem \ref{theorem:maximal-precise} constructs a
base $p$ expansion of $|\Omega(\II)|$ because each power of $p$ appears at most $p-1$ times.

\begin{corollary} \label{corollary:base-p-expansion}
Let $\II \subseteq [0:p^M-1]$. The formula \eqref{eq:Omega(I)-2} is of the form,
\[
|\Omega(\II)| = \sum_r p^{k_r} = \sum_s \alpha_s p^{l_s},
\]
with $l_1 > l_2 > l_3 > \ldots$ and $\alpha_s \in [0:p-1]$ for all $s$.
\end{corollary}

\begin{IEEEproof}  Begin with $\Omega(\II) = \bigcup_r\II_r^\dagger$.
Since the $\II_r^\dagger$ are disjoint, we have
\begin{equation}
\label{eq:uncen-phi-k-0}
\begin{aligned}
\sum_{r=1}^{\alpha_1} \widetilde{\chi}_{l_1+1}(a\,;\II_r^\dagger) &= \widetilde{\chi}_{l_1+1}(a\,;\bigcup_{r=1}^{\alpha_1}\II_r^\dagger) \nonumber\\
& \leq \widetilde{\chi}_{l_1+1}(a\,; \Omega(\II)), \quad a\in[0:p^{l_1+1}-1].
\end{aligned}
\end{equation}
Summing this over all $a \in [0:p^{l_1 +1}-1]$ we have 
\begin{equation}
\label{eq:uncen-phi-k}
\begin{aligned}
\alpha_1p^{l_1} &= \sum_{r=1}^{\alpha_1} |\II_r^\dagger| = \sum_{r=1}^{\alpha_1} \sum_a
\widetilde{\chi}_{l_1+1}(a\,;\II_r^\dagger)\\
& \leq \sum_i\widetilde{\chi}_{l_1+1}(a\,; \Omega(\II)) = |\Omega(\II)| < p^{l_1+1}, 
\end{aligned}
\end{equation}
so $\alpha_1 < p$. For the last inequality in
\eqref{eq:uncen-phi-k} we have used the upper bound from part (ii) in Theorem
\ref{thm:maximal-bounds}. We have also used that $|\II_r| = p^{k_r}$.
The proof for other $\alpha_s$ is similar. For example, to prove that
$\alpha_2<p$ we start with 
\[
\sum_{r=\alpha_1+1}^{\alpha_2} \widetilde{\chi}_{l_2+1}(a\,;\II_r^\dagger) \leq
\widetilde{\chi}_{l_2+1}\left(a\,; \Omega(\II_{a_1+1})\right)
\]
instead of \eqref{eq:uncen-phi-k-0}. 
 \end{IEEEproof}

If the algorithm above were initialized with a universal set $\II$,
then from Theorem \ref{theorem:maximal-precise} we would obtain
$\II= \Omega(\II)  =
\bigcup _ r \II_r^\dagger$. 
This allows us to conclude that any universal set $\II$ is a union of
elementary universal sets. 
Moreover, the sets $\II^\dagger_r$ defined by the algorithm satisfy
conditions in Lemma \ref{lem:univ-structure-forward}. For condition (ii), note that 
in the algorithm the set $\II_r$ is recursively defined as $ \II_r = \II_{r-1}\setminus
\mathcal{L}_{r-1}$, so that 
\[
\II_r = \left(\left(\left(\II \setminus
    \mathcal{L}_1\right)\setminus \mathcal{L}_2\right) \ldots
\setminus \mathcal{L}_{r-1}\right)= \II \setminus \left(
\bigcup_{j=1}^{r-1}\mathcal{L}_j \right).
\]
Hence $\II_r \cap
\left(\bigcup_{j=1}^{r-1} \mathcal{L}_j\right) = \emptyset$. Then the sets $\II^\dagger_r$, obtained by the algorithm, satisfy $\II_r^\dagger \cap
\left(\bigcup_{j=1}^{r-1} \mathcal{L}_j\right) = \emptyset$, since
$\II^\dagger_r \subseteq \II_r$. Putting all these comments together  we have the converse of
  Lemma \ref{lem:univ-structure-forward}, and then adding Theorem \ref{theorem:maximal-precise} we can state

\begin{corollary}
\label{cor:univ-structure-final}
 $\II \subseteq [0:p^M-1]$ is universal if and only if there
exist 
\begin{enumerate}[(i)]
\item A nonincreasing finite sequence $k_1 \ge k_2 \ge \cdots \ge 0$, with each value of $k_r$ repeating at most $p-1$ times;
\item Sets $\II^\dagger_r\subseteq \II$ with  $\II = \bigcup_r \II^\dagger_r$;
\end{enumerate}
such that 
\begin{enumerate}
\item[(iii)] $\II_r^\dagger$ is a $k_r$-elementary universal set;
\item[(iv)]  $\II_r^\dagger \cap \left(\bigcup_{j=1}^{r-1} \mathcal{L}_j\right)
  = \emptyset$, where
\[
\mathcal{L}_j=\{x \in [0:N-1]: x \equiv i \text{ mod $p^{k_j}+1$ for some $i \in
\II_j^\dagger$}\}. \nonumber
\]
\end{enumerate}
\end{corollary}

Note that from (i), (ii) and (iii) we can also
conclude that \[
|\II| = \sum_r |\II_r^\dagger| = \sum_r p^{k_r},
\]
so the $k_r$ are the powers of $p$ appearing in the base-$p$
expansion of $|\II|$, taken with repetitions.
For example with $N=9$, $|\II| = 7 = 2\cdot 3^1 + 1 \cdot 3^0$, we expect the
universal set $\II = \II_1^\dagger \cup \II_2^\dagger \cup
\II_3^\dagger$ with $\II_1^\dagger$ and $\II^\dagger_2$ being
$1$-elementary, and $\II_3^\dagger$ being $0$-elementary.
Corollary \ref{cor:univ-structure-final} implies that the $k_r$ read
off from the base-$p$ expansion of $|\II|$ must be the same as the
$k_r$ generated by the algorithm if $\II$ is universal.

\begin{remark}[Universal sets of prescribed order]
\label{rem:algo-mod}  As~it~stands, the  
algorithm finds a universal set of the largest size contained in $\II$. With Corollary \ref{cor:univ-structure-final} we can now modify the algorithm to solve the following problem:
\begin{quote}
\textit{Given  a set $\II \subseteq [0:p^M-1]$, and  
$d \le |\Omega(\II)|$, find a universal set $\JJ \subseteq \II$ with $|\JJ| = d$.}
\end{quote}
We follow the algorithm as in steps 1-4, but we change the definition of $k_r$ in Step 1. Write the base-p expansion of $d$ with repetitions, $d = \sum _r p^{k_r}$, read off the $k_r$ as the powers of $p$ that appear in the expansion, and arrange the $k_r$ in nonincreasing order. This ensures that condition (i) in Corollary \ref{cor:univ-structure-final} is satisfied. The construction of the $\II_r^\dagger$ in Steps 2-4 of the algorithm will ensure that (iii) and (iv) are satisfied. We conclude that with the $\II_r^\dagger$ so obtained by the algorithm the set $\JJ = \bigcup_r \II_r^\dagger \subset \II$ is universal, and it is of the right size by definition of the $k_r$.
\end{remark}

Finally, we have

\begin{IEEEproof}[Proof of Theorem \ref{theorem:maximal-precise}] As observed above, the $\II_r^\dagger$ generated by the algorithm satisfy
the hypotheses of Lemma \ref{lem:univ-structure-forward},  so the set
$\bigcup_r \II^\dagger_r$ is universal. If we show 
\[
|\Omega(I)| \leq \sum_rp^{k_r},
\]
then Theorem \ref{theorem:maximal-precise} follows.

For this we prove
\begin{equation} \label{eq:maximal-<=}
|\Omega(\II_r)| \le p^{k_r}+ |\Omega(\II_{r+1})|.
\end{equation}
We appeal to Theorem \ref{thm:maximal-bounds} to find a maximal universal sampling set $\A$ with $\II_r^{\dagger} \subseteq \A \subseteq \II_r$, and we will show
\begin{equation}
\A\setminus \II_r^\dagger \quad \text{is universal},\label{eq:last-condition-1} 
\end{equation}
\begin{equation}
\A\setminus \II_r^\dagger \subseteq \II_{r+1}. \label{eq:last-condition-2}
\end{equation}
Since $|\A| = |\Omega(\II_r)|$ These imply
\[
|\Omega(\II_r)| - p^{k_r} = |\A\setminus \II_r^\dagger| \le |\Omega(\II_{r+1})|,
\]
which is \eqref{eq:maximal-<=}.

First \eqref{eq:last-condition-1}. Now,
\[
\widetilde{\chi}_s(a \,;\A\setminus\II_r^\dagger) = \widetilde{\chi}_s(a \,;\A) - \widetilde{\chi}_s(a \,;\II_r^\dagger), \quad a \in[0:p^s-1],
\]
and for $s \le k_r$ the second term is constant,
\[
\widetilde{\chi}_s(a\,;\II_r^\dagger) = p^{k_r-s}, 
\]
from \eqref{eq:chi-constant}. Since $\A$ is universal, $|\widetilde{\chi}_{s}(a\,;\A) - \widetilde{\chi}_{s}(b\,;\A)| \le 1$ for all $a, b \in [0:p^s-1]$ and for all $s$, so we at least have
\[
|\widetilde{\chi}_s(a\,;\A\setminus\II_r^\dagger) - \widetilde{\chi}_s(b\,;\A\setminus\II_r^\dagger)| \le 1,
\]
for all $s \le k_r$. We need to check that this inequality continues to hold for $s \ge k_r+1$.

As we have argued before,  by the definition of $k_r$ at least one congruence class in $\II_r/p^{s}$ is empty when $s \ge k_r+1$, so $\widetilde{\chi}_s(a_0\,;\II_r)=0$ for some $a_0$, and because $\A \subseteq \II_r$ is universal we have $\widetilde{\chi}_s(a\,;\A) \le 1$ for all $a$. Furthermore,  $\I_r^\dagger \subseteq \A$ implies
\[
0 \le  \widetilde{\chi}_s(a\,;\A) - \widetilde{\chi}_s(a\,;\II_r^\dagger) = \widetilde{\chi}_s(a\,;\A\setminus\II_r^\dagger).
\]
Hence the values of
$\widetilde{\chi}_s(a\,;\A\setminus\II_r^\dagger)$ are in $\{0,1\}$ and consequently
\[
|\widetilde{\chi}_s(a\,;\A\setminus\II_r^\dagger)-\widetilde{\chi}_s(b\,;\A\setminus\II_r^\dagger)| \le 1,
\]
for all $s \ge k_r+1$. This establishes that $\A\setminus \II_r^\dagger$ is universal.

We prove \eqref{eq:last-condition-2} by contradiction.  If it were not true that $\A\setminus \II_r^\dagger \subseteq \II_{r+1}$ then there would exist an $x \in (\A\setminus \II_r^\dagger) \cap \LL_r$. Then $\widetilde{\chi}_{k_r+1}([x]_{k_r+1}\,;\A) = 2$, for on dividing by $p^{k_r+1}$,  $x$ leaves a remainder of $[x]_{k_r+1}$ (by definition) and so does one other element in $\II_r^\dagger$. But this contradicts  $\widetilde{\chi}_s(a\,;\A) \le 1$ for $s \ge k_r+1$ from the preceding paragraph. 

This completes the proof of Theorem  \ref{theorem:maximal-precise}. 
\end{IEEEproof}

\begin{remark}
We can give an upper bound for the computational complexity of the algorithm for constructing a universal sampling set of size $d$ (including constructing a maximal universal sampling set). Within an iteration, in the worst case the algorithm makes a complete
pass over all the nodes of the tree once, and the the number of nodes is $O(N)$. Further, the number of iterations is $\alpha_1+\alpha_2+\cdots +\alpha_M$ where
\[
d = \alpha_1p^{M-1} + \alpha_{2}p^{M-2} + \ldots + \alpha_{M-1}p + \alpha_M.
\]
Hence the largest number of iterations is $(p-1)M$, and the complexity
of the algorithm is at most $O(N \log N)$.
\end{remark}

\subsection{Counting Universal Sets} \label{subsection:counting-universal}

The preceding structure theorems 
  allow us to find the number of universal sampling sets $\II \subseteq [0:p^M-1]$ of size $d$. The formula uses the digits from the   base-$p$ expansion of $d$, and as above we let
\begin{equation}
\nonumber
d = \alpha_1p^{M-1} + \alpha_{2}p^{M-2} + \ldots + \alpha_{M-1}p + \alpha_M,
\end{equation}
where $0 \leq \alpha_i < p$. For $i=0,1,\dots, M$ define
\[
d_i= \sum_{j=i+1}^M\alpha_jp^{M-j}.
\]
Hence $d_0=d$ and $d_M=0$.

\begin{theorem} \label{thm:count}
The number of universal sampling sets in $[0:p^M-1]$ of size $d$ is 
\[
\mathcal{C}(d,p^M) = \prod_{i=1}^M \binom{p}{\alpha_i+1}^{d_i}\binom{p}{\alpha_i}^{p^{M-i}-d_i} .
\]
\end{theorem}

\begin{IEEEproof} The proof goes by establishing a recurrence relation for $\mathcal{C}$ in the $d_i$.\footnote{We are grateful to a reviewer for suggesting a way to make greater use of the recursive aspect of our original argument, resulting in a much shorter and cleaner proof.} Let $\II$ be a universal sampling set of size $d$ and construct the congruence tree as in Figure \ref{fig:tree1} with root $\II_{00}=\II$. We first note that $d_1$ of the nodes at level $M-1$ have weight $\alpha_1+1$ and the remaining $p^{M-1}-d_1$ nodes have weight $\alpha_1$, where
\[
d_1 = \sum_{i=2}^M \alpha_i p^{M-i}.
\]
 The proof for this is along the same lines as the argument in the proof of Theorem \ref{theorem:universal}, $(i) \Longleftrightarrow (ii)$. 
Figure \ref{fig:tree10} illustrates this. The singleton blue nodes at the bottom level are the elements of $\II$, and the other nodes (which would be the singletons $\{6\}$ and $\{7\}$) are empty. The red nodes at the penultimate
level represent the nodes that have weight $\alpha_1 +1$ (and there
are $d_1$ of them). 

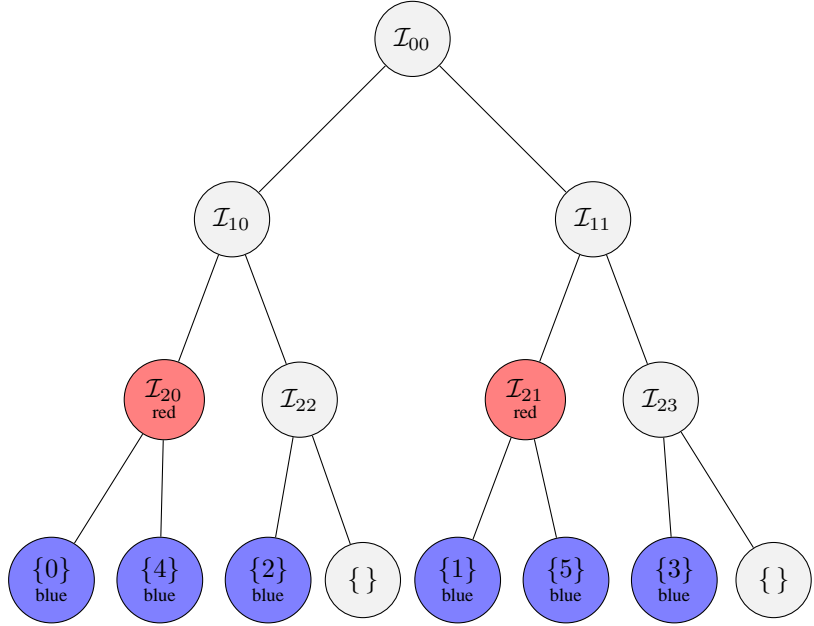
\begin{figure*}
\centering
\begin{tikzpicture}[scale=1.2]
\node (0) at (0,8) [circle, minimum size = 1.0cm, fill=gray!10, draw] {$\II_{00}$};
\node (1) at (-2,6) [circle, minimum size = 1.0cm, fill=gray!10,
draw] {$\II_{10}$};
\node (2) at (2,6) [circle, minimum size = 1.0cm, fill=gray!10, draw]
{$\II_{11}$};
\node (3) at (-2.75,4) [circle, minimum size = 1.0cm, fill=red!50, draw]
{$\underset{\text{red}}{\II_{20}}$};
\node (4) at (-1.25,4) [circle, minimum size = 1cm, fill=gray!10, draw]
{$\II_{22}$};
\node (5) at (1.25,4) [circle, minimum size = 1cm, fill=red!50, draw]
{$\underset{\text{red}}{\II_{21}}$};
\node (6) at (2.75,4) [circle, minimum size = 1cm, fill=gray!10, draw]
{$\II_{23}$};
\node (7) at (-4,2) [circle, minimum size = 1cm, fill=blue!50, draw]
{$\underset{\text{blue}}{\{0\}}$};
\node (8) at (-2.8,2) [circle, minimum size = 1cm, fill=blue!50, draw]
{$\underset{\text{blue}}{\{4\}}$};
\node (9) at (-1.6,2) [circle, minimum size = 1cm, fill=blue!50, draw]
{$\underset{\text{blue}}{\{2\}}$};
\node (10) at (-0.55,2) [circle, minimum size = 1cm, fill=gray!10, draw]
{$\{\,\}$};
\node (11) at (0.5,2) [circle, minimum size = 1cm, fill=blue!50, draw]
{$\underset{\text{blue}}{\{1\}}$};
\node (12) at (1.7,2) [circle, minimum size = 1cm, fill=blue!50, draw]
{$\underset{\text{blue}}{\{5\}}$};
\node (13) at (2.9,2) [circle, minimum size = 1cm, fill=blue!50, draw]
{$\underset{\text{blue}}{\{3\}}$};
\node (14) at (4,2) [circle, minimum size = 1cm, fill=gray!10, draw]
{$\{\,\}$};
\draw (node cs:name=0) -- (node cs:name =1);
\draw (node cs:name=0) -- (node cs:name =2);
\draw (node cs:name=1) -- (node cs:name =3);
\draw (node cs:name=1) -- (node cs:name =4);
\draw (node cs:name=2) -- (node cs:name =5);
\draw (node cs:name=2) -- (node cs:name =6);
\draw (node cs:name=3) -- (node cs:name =7);
\draw (node cs:name=3) -- (node cs:name =8);
\draw (node cs:name=4) -- (node cs:name =9);
\draw (node cs:name=4) -- (node cs:name =10);
\draw (node cs:name=5) -- (node cs:name =11);
\draw (node cs:name=5) -- (node cs:name =12);
\draw (node cs:name=6) -- (node cs:name =13);
\draw (node cs:name=6) -- (node cs:name =14);
\end{tikzpicture}
\caption{The congruence class tree for $N=8$. The universal sampling set $\{0,1,2,3,4,5\}$ of size
  $d=6$ is represented by the blue nodes at the bottom level. The
  red nodes at the penultimate level represent the nodes that
have weight 2, the rest of the nodes at the penultimate level have
weight 1.}
\label{fig:tree10}
\end{figure*}

Now remove the bottom level of the tree, effectively making
$N=p^{M-1}$, and resulting in Figure \ref{fig:tree11}. If the starting set (the blue nodes in
Figure \ref{fig:tree10}) is universal, then so must be the set formed
by the red nodes in Figure \ref{fig:tree11}. Hence the number of ways of
choosing the red nodes is the same as the number of universal sampling sets of
size $d_1$ in $[0:p^{M-1}-1]$, that is $\mathcal{C}(d_1,p^{M-1})$..

Once the red nodes are chosen, we need to choose the blue nodes by taking
$\alpha_1+1$ elements from the red nodes and $\alpha_1$ elements from
the remaining (non-red) nodes, which can be done in
\[
\binom{p}{\alpha_1+1}^{d_1}\binom{p}{\alpha_1}^{p^{M-1}-d_1}
\]
ways. Hence
\begin{equation} \label{eq:count-recurrence}
\mathcal{C}(d, p^M) = \binom{p}{\alpha_1+1}^{d_1}\binom{p}{\alpha_1}^{p^{M-1}-d_1}\mathcal{C}(d_1, p^{M-1}).
\end{equation}
This full formula follows.
\end{IEEEproof}

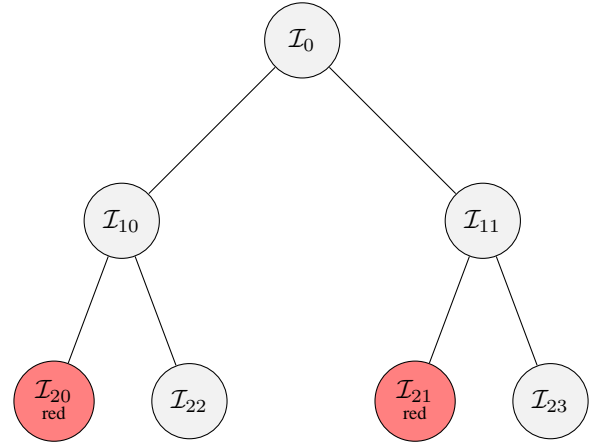
\begin{figure}
\centering
\begin{tikzpicture}[scale=1.2]
\node (0) at (0,8) [circle, minimum size = 1.0cm, fill=gray!10, draw] {$\II_0$};
\node (1) at (-2,6) [circle, minimum size = 1.0cm, fill=gray!10,
draw] {$\II_{10}$};
\node (2) at (2,6) [circle, minimum size = 1.0cm, fill=gray!10, draw]
{$\II_{11}$};
\node (3) at (-2.75,4) [circle, minimum size = 1.0cm, fill=red!50, draw]
{$\underset{\text{red}}{\II_{20}}$};
\node (4) at (-1.25,4) [circle, minimum size = 1cm, fill=gray!10, draw]
{$\II_{22}$};
\node (5) at (1.25,4) [circle, minimum size = 1cm, fill=red!50, draw]
{$\underset{\text{red}}{\II_{21}}$};
\node (6) at (2.75,4) [circle, minimum size = 1cm, fill=gray!10, draw]
{$\II_{23}$};
\draw (node cs:name=0) -- (node cs:name =1);
\draw (node cs:name=0) -- (node cs:name =2);
\draw (node cs:name=1) -- (node cs:name =3);
\draw (node cs:name=1) -- (node cs:name =4);
\draw (node cs:name=2) -- (node cs:name =5);
\draw (node cs:name=2) -- (node cs:name =6);

\end{tikzpicture}
\caption{Remove the bottom level of the tree in Figure \ref{fig:tree10}. The resulting red nodes
  are a universal sampling set in $[0:3]$}
  \label{fig:tree11}
\end{figure}

One special case of the counting formula is easy to evaluate.

\begin{corollary}
Let $d = p^k$ where $k <M$. Then the number of universal sets of size $d$ in $[0:p^M-1]$ is $(p^M/d)^d$.  
\end{corollary}

In particular when $N=2^M$, and $d = 2^{M-1}=N/2$, the number of universal sets is $2^{N/2}$.  
On the other hand, the total number of sets of size $2^{M-1}$ in $[0:2^N-1]$ is 
\[
\binom{N}{N/2} \approx 2^N/\sqrt{\pi N}
\] 
by Stirling's approximation. Hence the fraction of sets that are universal is approximately $\sqrt{\pi N}/2^{N/2}$, which decreases exponentially with $N$.

The function $\mathcal{C}(d,p^M)$ is certainly complicated, but 
it has some remarkable properties.  Though not clear from the formula, we have 
\[
\mathcal{C}(d,p^M) = \mathcal{C}(p^M-d,p^M).
\]
 This follows from the following lemma, which is itself a simple but interesting property of universal sampling sets.

\begin{lemma} \label{lemma:complement-universal-is-universal}
 If $\A \subseteq[0:p^M-1]$ is a universal sampling set then so is $\A'=[0:p^M-1] \setminus \A$.
\end{lemma}

This extends the bracelet property of universal sampling sets, though for bracelets we need not assume that $N$ is a prime power. 

\begin{IEEEproof}
For any $0 \le k \le M$ and $a \in [0:p^k-1]$,
\[
\begin{aligned}
\widetilde{\chi}_k(a\,;\A') &= \widetilde{\chi}_k(a\,;[0:p^M-1]) - \widetilde{\chi}_k(a\,;\A)\\
&= p^{M-k} - \widetilde{\chi}_k(a\,;\A).
\end{aligned}
\]
 
Next, since 
\[
|\widetilde{\chi}_k(a\,;\A) - \widetilde{\chi}_k(b\,;\A)| \leq 1,
\]
for all $a,b \in [0:p^k-1]$, it follows that 
\[
|\widetilde{\chi}_k(a\,;\A') - \widetilde{\chi}_k(b\,;\A')| \leq 1 .
\]
\end{IEEEproof}

\begin{figure*}
\centering
\includegraphics[width =\textwidth]{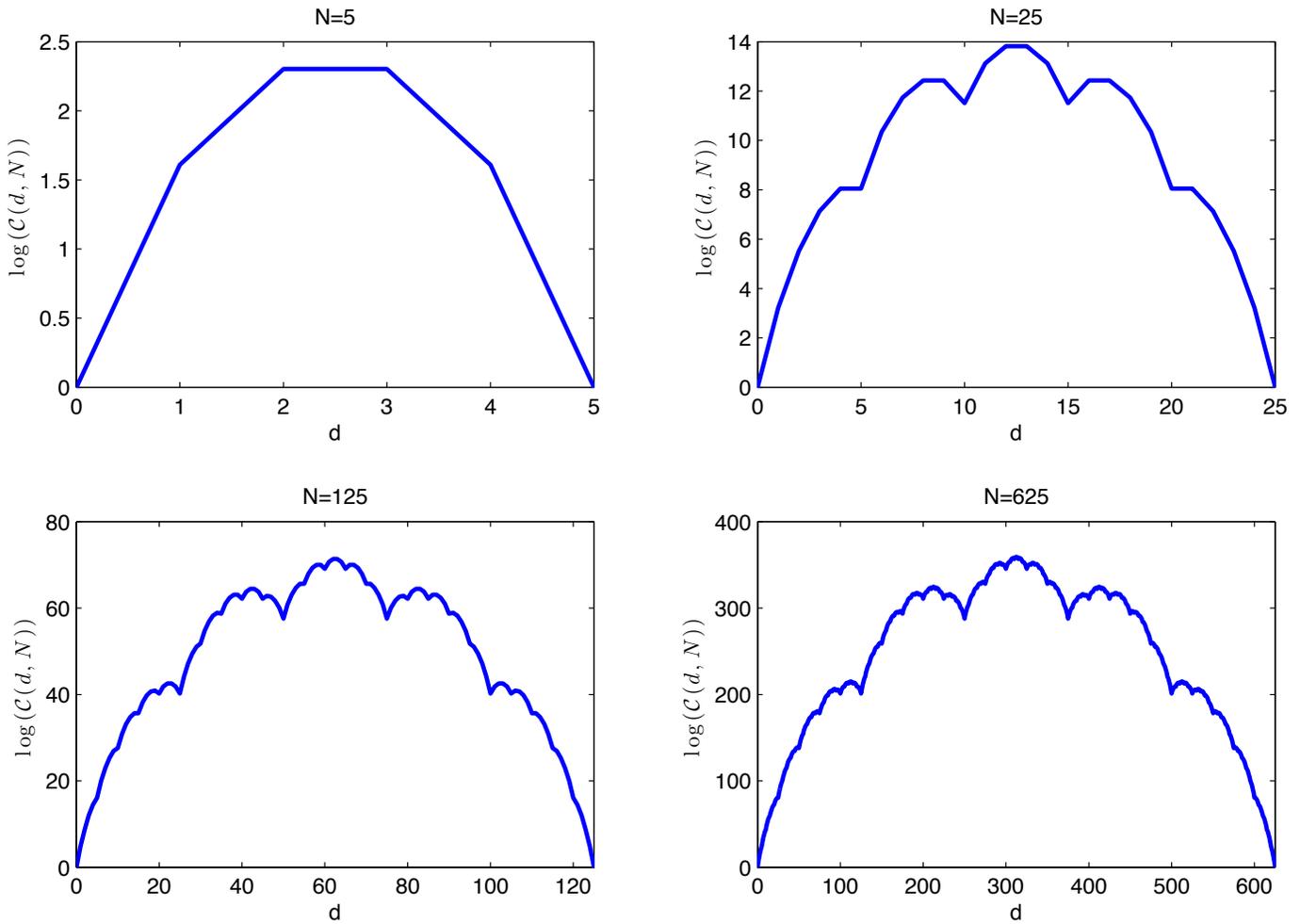}
\caption{Plots of $\log\mathcal{C}(d,p^M)$ \textit{vs} $d$ for powers of $p=5$. Note the $5$ bumps on different scales.}
\label{fig:count}
\end{figure*}

Figure \ref{fig:count} displays $\log \mathcal{C}(d,5^M)$ as a function of $d$ as $M$ takes
increasing values. The plots show the symmetry, $\mathcal{C}(d,p^M) = \mathcal{C}(p^M-d,p^M)$, but they show much more. We can observe the following:
\begin{enumerate}[(i)]
\item  There are a series of bumps on several (visible) scales. One
  cannot fail to notice that at each scale the number of bumps in the
  graph is $5$, which is the prime $p$ here. Experiments with other
  primes have similar plots and in each case indicate   that the number of bumps is equal to the prime. 
  \end{enumerate}

\begin{enumerate}
\item[(ii)] With increasing $M$ the plots of the
  count are somehow converging in shape -- they all start to look
  similar. 
 \end{enumerate}
  The second point can indeed be quantified. One can show that for each 
  $\alpha \in [0,1]$, 
\[
\lim_{M \rightarrow \infty}\frac{\log \mathcal{C}(\lfloor\alpha p^M\rfloor,p^M)}{p^M}
\]
exists. See \cite{Aditya:thesis}.  This compares nicely with the fact that a similar
function with $\mathcal{C}(d,N)$ replaced by $\binom{N}{d}$  also converges, and to the entropy function:
\[
\begin{aligned}
\lim_{M \rightarrow \infty}\left(\frac{1}{p^M}\log \binom{p^M}{\lfloor\alpha  p^M\rfloor}\right) &= \alpha \log \frac{1}{\alpha} + (1-\alpha)\log\frac{1}{1-\alpha}\\
& =H(\alpha).
\end{aligned}
\]
This is the limiting case of counting \emph{all} index sets.

\begin{figure*}
\centering
\includegraphics[width = .9\textwidth]{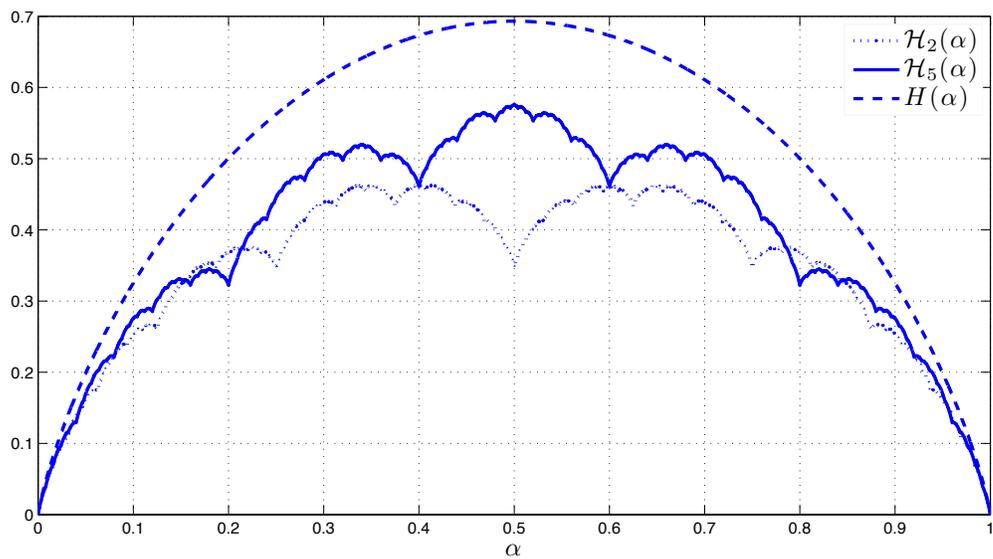}
\caption{Plots of the limit of the counting functions for $p=2,5$ compared to the Entropy function. Note the self-similarity as it depends on the prime.}
\label{fig:count-entropy}
\end{figure*}

\begin{figure*}
\centering
\includegraphics[width = .9\textwidth]{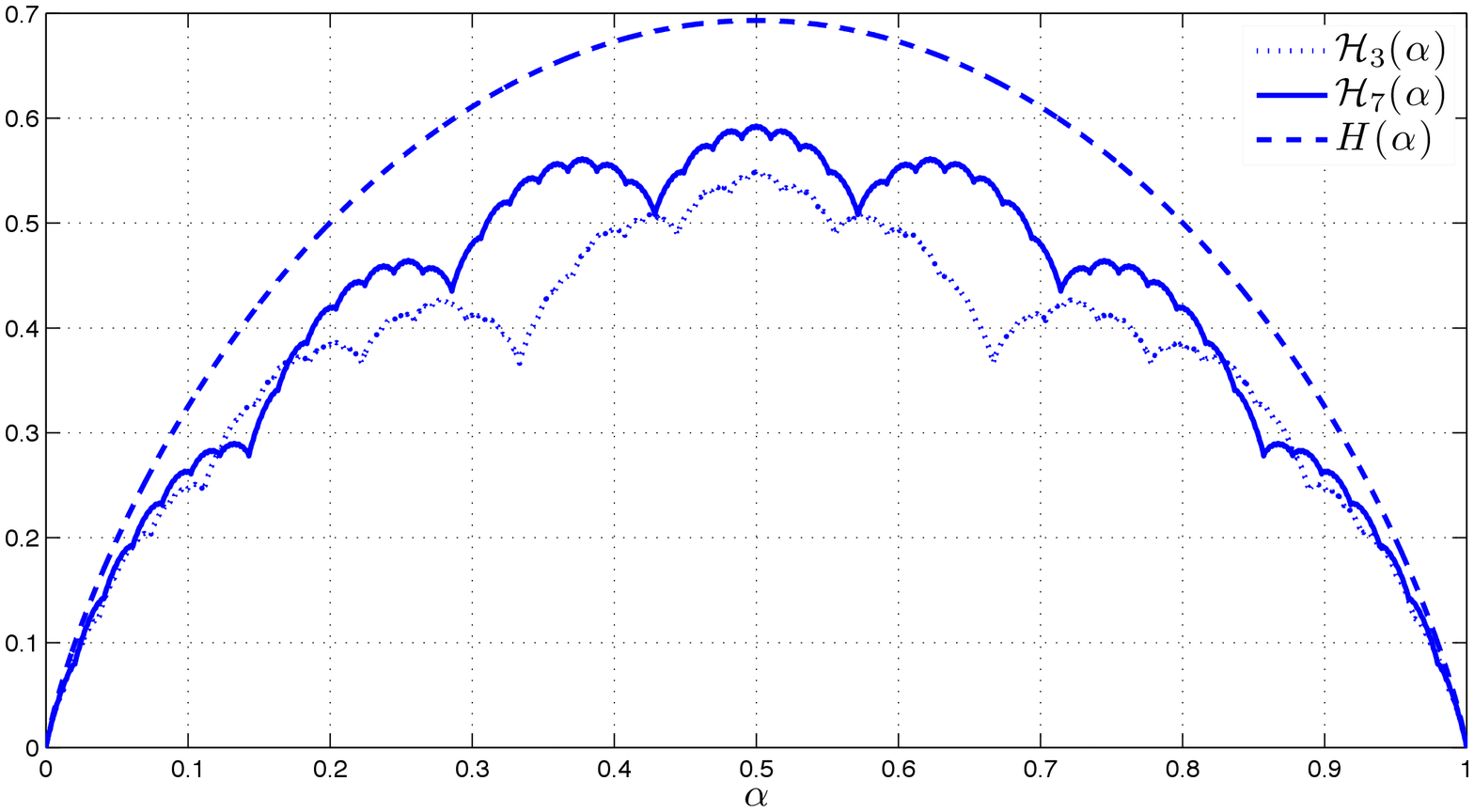}
\label{fig:count-converging-1}
\caption{Similar to Figure \ref{fig:count-entropy} with $p=3,7$}
\label{fig:count-converging}
\end{figure*}

Plots of
\begin{equation} \label{eq:count-limit}
\mathcal{H}_p(\alpha) = \lim_{M \rightarrow \infty}\frac{\log \mathcal{C}(\lfloor\alpha p^M\rfloor,p^M)}{p^M}, \quad 0\le \alpha \le 1,
\end{equation}
are shown in Figures \ref{fig:count-entropy} and
\ref{fig:count-converging} for several values of $p$, along with a plot of $H(\alpha)$.


The plots of $H_p(\alpha)$ seem to satisfy observation (i), that the curves have $p$ bumps at each scale.  Here is an explanation. In the notation of Theorem \ref{thm:count}, suppose $\alpha_1=0$ (i.e., $d <p^{M-1}$). Then $d_1=d$ and we have, as in \eqref{eq:count-recurrence},
\[
\begin{aligned}
\mathcal{C}(d,p^M) &= \binom{p}{1}^{d_1}\binom{p}{0}^{p^{M-1}-d_1}\mathcal{C}(d_1,p^{M-1})\\
& = p^d\mathcal{C}(d,p^{M-1}).
\end{aligned}
\]
Let $M \rightarrow \infty$, so $d/p^M \rightarrow \alpha$ (with $\alpha < p$). Then with reference to \eqref{eq:count-limit}, 
\[
\mathcal{H}_p(\alpha) = \lim_{M\rightarrow \infty}\left(\frac{d}{p^M}\log p + \mathcal{H}_p(p\alpha)\right) = \alpha \log p +\mathcal{H}_p(p\alpha),
\]
leading to the self-similar plots we observe.

\subsection{Maximal and Minimal Universal Sampling Sets} \label{subsection:maximal-minimal}

Along with maximal universal sets is the allied notion of  minimal universal sets.

\begin{defn} \label{definition:minimal}
Let $\II \subseteq [0:N-1]$. A \emph{minimal universal sampling set} for $\II$ is a universal sampling set of smallest cardinality that contains $\II$.
\end{defn}

Again we need a notation and  we let $\Phi(\II)$ denote a generic minimal universal sampling set containing $\II$. Thus $|\Phi(\II)| \le |\JJ|$ for any universal sampling set $\JJ \supseteq \II$.

Let us show one way that maximal and minimal universal sampling sets are related. The proof relies on Lemma \ref{lemma:complement-universal-is-universal} from the previous subsection.

\begin{theorem} \label{theorem:maximal-minimal}
Let $\II \subset [0:p^M-1]$, $\II'=[0:p^M-1]\setminus \II$. Then
\[
|\Phi(\II)| = p^M-|\Omega(\II')|.
\]
\end{theorem}

\begin{IEEEproof}
Let $\A'=[0:p^M-1]\setminus \Phi(\I)$. Then $\A'$ is universal by Lemma \ref{lemma:complement-universal-is-universal}. Since $\Phi(\II) \supseteq \II$ we have $\A' \subset [0:N-1] \setminus \II =\II'$ and hence
\[
p^M-|\Phi(\II)| = |\A'| \le |\Omega(\II')|.
\]

Similarly, let $\B'=[0:p^M-1]\setminus \Omega(\II')$. Then $\B'$ is universal, it contains $[0:p^M-1]\setminus \II'=\II$ and so
\[
p^M-|\Omega(\II')| = |\B'| \ge |\Phi(\II)|.
\]
Taken together the two inequalities prove the theorem.
\end{IEEEproof}

\section{An Uncertainty Principle, Random Signals, and Sumsets} \label{section:uncertainty}

Generally speaking, an ``uncertainty principle'' is an inequality relating the supports of a nonzero function and its Fourier transform, in the present setting  $f\colon \mathbb{Z}_N \longrightarrow \mathbb{C}$,  and $\uFT f\colon \mathbb{Z}_N \longrightarrow \mathbb{C}$.  The notions of  maximal and minimal universal sampling sets lead immediately to an additive uncertainty principle. Without the language of universality, Tao \cite{tao:uncertainty} made this connection in the case when $N$ is a prime using Chebotarev's theorem, see Corollary \ref{corollary:uncertainty-prime}, though, as he states, it was probably already known as a folk theorem.

Let 
\[
\mathcal{Z}(f) = \{i \colon f(i) = 0\} 
\]
be the zero set of $f$. The support is the complement of the zero set, and we denote it by $\supp(f)$. Our result is
\begin{theorem} \label{theorem:uncertainty}
If $f$ is not the zero function then
\begin{equation} \label{eq:uncertainty-1}
\begin{aligned}
|\supp(\uFT f)| &\ge 1+|\Omega(\mathcal{Z}(f))|, \; \\
|\supp (f)| &\ge 1+|\Omega(\mathcal{Z}(\uFT f))|;
\end{aligned}
\end{equation}
and
\begin{equation} \label{eq:uncertainty-2}
\begin{aligned}
|\mathcal{Z}(\uFT f)| +1 &\le |\Phi(\supp(f))|,\; \\
 |\mathcal{Z}( f)| +1 &\le |\Phi(\supp(\uFT f))|.
 \end{aligned}
\end{equation}
\end{theorem}

We are not assuming that $N$ is a prime power here. However, we immediately deduce
\begin{corollary}[Tao]  \label{corollary:uncertainty-prime}
If $N$ is prime and $f$ is not the zero function then 
\[
|\supp(\uFT{f})|  + |\supp (f)| \geq N +1.
\]
\end{corollary}
\begin{IEEEproof} If $N$ is prime, then by Chebotarev's theorem every index set
is universal. In particular the set $\mathcal{Z}(f)$ is universal. Hence
$\Omega(\mathcal{Z}(f)) = \mathcal{Z}(f)$. From Theorem \ref{theorem:uncertainty},
 \[
 \begin{aligned}
|\supp(\uFT {f})| &\geq 1 + |\Omega(\mathcal{Z}(f))| \\
&= 1 + |\mathcal{Z}(f)| = 1 + N - |\supp(f)|.
\end{aligned}
\]
\end{IEEEproof}

We also have

\begin{corollary} \label{corollary:uncertainty-consecutive}
Suppose $f$ vanishes on a set of consecutive integers $\II$. Then
$|\supp(\uFT{f})| \geq |\II|+1$. If $\JJ$ is a set of
integers such that $\uFT f(\JJ) = 0$, then $|\II| + |\JJ| \leq N - 1$.
\end{corollary}

\begin{IEEEproof}We observed previously that any set of consecutive integers, $\II$ in this case, is
universal. Since $\II \subseteq \mathcal{Z}(f)$, we have $|\Omega(\mathcal{Z}(f))| \geq |\II|$. From Theorem
\ref{theorem:uncertainty}, this implies $|\supp(\uFT{f})| \geq
|\II| + 1$. Further, if $\uFT{f}(\JJ) = 0$ then $N - |\JJ| \geq |\supp (\uFT f)|$ and
so  $N - |\JJ| \geq  |\II| + 1$.
\end{IEEEproof}

The proof of Theorem \ref{theorem:uncertainty} itself is very brief. 
\begin{IEEEproof}[Proof of Theorem \ref{theorem:uncertainty}] Suppose
$|\supp (\uFT{f})| \leq |\Omega(\mathcal{Z}(f))|$. From
$\Omega(\mathcal{Z}(f)) \subseteq \mathcal{Z}(f)$ it follows that $f$ vanishes on $\Omega(\mathcal{Z}(f))$. Since $\Omega(Z)$ is a universal sampling set this implies that $\uFT{f}
\equiv 0$,  contradicting the assumption that $f$ is not the zero function. This proves the first statement in \eqref{eq:uncertainty-1}.  A similar argument establishes the second statement.

For the proof of \eqref{eq:uncertainty-2}, write $\mathcal{Z} = \mathcal{Z}(\mathcal{F}f)$ and $\A=\Phi(\text{supp}(f))$. Then
\[
\mathcal{F}f (\ZZ) = 0 \text{ and so } E_{\ZZ}^\textsf{T}\mathcal{F}f  = 0.
\]
However $f$ is supported within $\A$, and so we may
write $f = E_{\A}g$, where $ g = f(\A) \neq 0$. This means we must have
\begin{equation}
\label{eq:alt-proof}
 E_{\ZZ}^\textsf{T}\mathcal{F}E_{\A} g = 0, \text{ for  some } g \neq 0, 
\end{equation}
i.e. the columns of  $E_{\ZZ}^\textsf{T}\mathcal{F}E_{\A}$ are
dependent. This is expected if $|\ZZ|< |\A|$. However, if $|\ZZ| \geq |\A|$,
this contradicts the universality of $\A$. Hence we
must have $|\ZZ| \leq |\A|-1$, which is the first inequality in \eqref{eq:uncertainty-2}. A similar argument establishes the second statement.
\end{IEEEproof}

It is interesting that when $N$ \emph{is} a prime power the two statements \eqref{eq:uncertainty-1} and \eqref{eq:uncertainty-2} are equivalent. To see this we first derive \eqref{eq:uncertainty-2} from \eqref{eq:uncertainty-1} when $N=p^M$. This appeals to Theorem \ref{theorem:maximal-minimal} on the relation between maximal and minimal sets, with $\supp(f) = [0:N-1]\setminus \mathcal{Z}(f)$. Thus, from \eqref{eq:uncertainty-1}, $|\supp(\uFT f)| \ge 1+|\Omega(\mathcal{Z}(f))|$, and substituting from Theorem \ref{theorem:maximal-minimal},
\[
|\supp(\uFT f)| \ge 1 + N - |\Phi(\supp(f))|.
\]
But $|\supp(\uFT f)| = N-|\mathcal{Z}(\uFT f)|$, so
\[
N-|\mathcal{Z}(\uFT f)| \ge 1+N-|\Phi(\supp(f))|,
\]
which is the same as the first statement in \eqref{eq:uncertainty-2}. Again, the second statement in \eqref{eq:uncertainty-2} follows in a similar manner. We could have started instead with \eqref{eq:uncertainty-2} and from this derived \eqref{eq:uncertainty-1}.

In cases where $\mathcal{Z}(f)$ itself is a universal sampling set, the uncertainty
principle in Theorem \ref{theorem:uncertainty} can be as strong as the uncertainty
principle for the prime $N$ case.

\begin{remark}
Readers familiar with the seminal paper of Donoho and Stark \cite{donoho-stark:uncertainty} will wonder if the additive uncertainty principle in Theorem \ref{theorem:uncertainty} can be applied to the problem of reconstruction of a signal corrupted
by sparse noise. (See also \cite{studer-signal-recovery} for more recent work.)  The answer is yes, and we refer to \cite{osw:reconstruction}.

\end{remark}

\subsection{Random Index Sets and Random Signals}

We will give several applications of these ideas. First we combine Theorem \ref{theorem:uncertainty} with a probabilistic estimate on the size of a maximal universal sampling set for randomly chosen index sets. We must revert to the assumption that $N$ is a prime power.

\begin{theorem}
\label{theorem:rand-samp}
Let $N=p^M$. Let $\mathcal{R}_s$ be an index set of $s$ numbers chosen at random from $[0:N-1]$. Let $\lambda = (N-s)/N$. If $d, \delta>0$ satisfy
\begin{equation}
\label{eq:ran-uncen-1}
N\log(1/\lambda) \geq (1+ \delta) d \log d, 
\end{equation}
then $|\Omega(\mathcal{R}_s)| \geq d$ with probability at least $1- d^{-\delta}$.
\end{theorem}

This means that if we can choose a large $d$ satisfying \eqref{eq:ran-uncen-1}, which is possible, for example, if $N$ is large and $\lambda$ is small, then $|\Omega(\mathcal{R}_s)| \geq d$ with high probability. Thus while it is unlikely that a randomly chosen index set will be universal, it is quite likely that such an index set will contain a large universal set as a subset.

We will apply Theorem \ref{theorem:rand-samp} to the case when $\mathcal{R}_s$ is the zero set of $f\colon \mathbb{Z}_N \longrightarrow \mathbb{C}$. Then $\lambda = |\text{supp}(f)|/N$, i.e., $\lambda$ is the fraction of nonzero entries in $f$.

\begin{IEEEproof} The proof uses the bound in part (ii) of Theorem \ref{thm:maximal-bounds}. Let $k$ be the largest integer such that no congruence classes in $\RR_s/p^k$ are empty. Note that $k$ is random since $\RR_s$ is random. Then $|\Omega(\RR_s)| \le d-1$ implies 
\[
p^k \le |\Omega(\RR_s)| \le d-1,
\]
by Theorem \ref{thm:maximal-bounds}. Therefore
\begin{align}
&\text{Prob}\left(|\Omega(\RR_s)| \leq d-1 \right) \leq \text{Prob}(p^k \leq d-1) \nonumber \\
&\quad = \text{Prob}(k \leq \lfloor \log_p(d-1) \rfloor) \nonumber \\
&\quad = \text{Prob}(\text{at least one congruence class in } \nonumber\\
& \hspace{.5in} \RR_s/p^{\lfloor \log_p(d-1) \rfloor +1} \text{ is empty}).
\end{align}
We will compute the last probability.

Let $b = {\lfloor \log_p(d-1) \rfloor +1}$, and let $\mathcal{N}_{ba}$ be the set of elements in $[0:N-1]$ that leave a remainder of $a\in[0:p^b-1]$ when divided by $p^b$.  Since $N=p^M$ all of the $\mathcal{N}_{ba}$ have size $t = N/p^b  = p^M / p^{\lfloor \log_p(d-1) \rfloor +1}$.  

Fix a particular residue $a$. The probability that $\mathcal{N}_{ba}\cap \RR_s $ is empty (in words, the probability that a particular congruence class goes missing in $\RR_s$) is  $\binom{N-t}{s}/\binom{N}{s}$. This is because the number of ways of picking $\RR_s$ is $\binom{N}{s}$ while the number of ways of picking $\RR_s $ so that $\mathcal{N}_{ba} \cap \RR_s = \emptyset$ is the number of ways of picking $s$ elements from 
\[ |[0:N-1] \setminus \mathcal{N}_{ba}| = N- t
\]
 elements. Then
\begin{align}
&\text{Prob}\left(\mathcal{N}_{ba}\cap \RR_s = \emptyset\right) \nonumber \\ 
&= \binom{N-t}{s}\Big{/} \binom{N}{s} \nonumber \\
&= \frac{(N-(t-1)
  -s)(N-(t-2)-s)\ldots(N-s)}{(N-t+1)(N-t+2)\ldots N} \nonumber \\
&\quad = \left(1- \frac{s}{N-t+1} \right) \left(1- \frac{s}{N-t+2}
\right)\ldots \left(1- \frac{s}{N} \right) \nonumber \\
&\quad \leq \left(1- \frac{s}{N} \right) \left(1- \frac{s}{N}
\right)\ldots \left(1- \frac{s}{N} \right) =
\left(1-\frac{s}{N}\right)^t. \label{eq:rnd-1} 
\end{align}
From this,
\begin{align}
 &\text{Prob}(\text{at least one congruence class in
   } \nonumber \\
 & \hspace{.7in} \RR_s/p^{\lfloor \log_p(d-1) \rfloor +1} \text{ is empty})
 \nonumber \\
 &\quad = \text{Prob}\left(\bigcup_i \left(\mathcal{N}_{ba}\cap \RR_s = \emptyset \right)\right)
 \nonumber \\
&\quad \leq \sum_i  \text{Prob}\left(\mathcal{N}_{ba}\cap \RR_s = \emptyset\right) \nonumber \\
&\quad \leq  \frac{N}{t}\left(1-\frac{s}{N}\right)^t = N\lambda^t/t.\label{eq:end-2}
\end{align}
Hence we have from \eqref{eq:end-2},
\begin{equation}
\label{eq:ran-bnd-1}
\text{Prob}\left(|\Omega(\RR_s)| 
\leq d-1 \right) \leq N\lambda^t/t. 
\end{equation}
Now,  $ t =  N / p^{\lfloor \log_p(d-1) \rfloor +1} \geq N/d$, since $\lfloor x \rfloor \leq x$. Using this in \eqref{eq:ran-bnd-1},
\begin{align}
&\text{Prob}\left(|\Omega(\RR_s)| \leq d-1 \right) \nonumber\\ 
& \leq  N\lambda^t/t \leq d\lambda^{N/d} \nonumber \\
& =\exp\left(\log d - \frac{N\log(1/\lambda)}{d} \right) \nonumber \\
& = \exp\left(\log d\left(1- \frac{N\log(1/\lambda)}{d\log d} \right)\right) \nonumber \\
& \leq \exp\left(-\delta \log d\right) \text{ (from the hypothesis of the theorem) } \nonumber \\
& = d^{-\delta}.
\end{align}
We conclude that $\text{Prob}\left(|\Omega(\RR_s)| \geq d \right) \geq 1- d^{-\delta}$.
\end{IEEEproof}

 We can now state a probabilistic uncertainty principle. Afterward we will comment on how this compares to the result of Candes, Romberg and Tao \cite{candes:robust}. 
 
 \begin{theorem}
\label{thm:rnd-uncen} 
Let $N=p^M$. Let $\mathcal{G}_{N,r}$ be the set of all signals $g\colon \mathbb{Z}_N \longrightarrow \mathbb{C}$ with support of size $r$. Let $g \in \mathcal{G}_{N,r}$ be a signal whose support is drawn at random from the set of all index sets of size $r$. Let the values of $g$ on the support set be drawn according to some arbitrary distribution. For $\delta>0$ let 
\[
a_{N, \delta} = \frac{N}{(1+\delta)\log N}\left(1+ \log (1+ \delta) + \log \log N\right).
\]
Then 
\begin{equation}
\label{eq:rnd-uncen}
|\text{supp}(g)| + |\text{supp}(\mathcal{F}g)| \geq 1 + a_{N,\delta} 
\end{equation}
with probability at least $1- (a_{N,\delta}-r)^{-\delta}$.
\end{theorem}

If $r$ is small compared to $a_{N,\delta}$, Theorem \ref{thm:rnd-uncen} states
that almost all signals $g$ in $\mathcal{G}_{N,r}$ satisfy the 
uncertainty principle above; roughly speaking
\[
|\text{supp}(g)| + |\text{supp}(\mathcal{F}g)| \geq N(1+ \log \log
N)/\log N
\]
for most $g$.
\begin{IEEEproof}
Picking the support of $g$  at random among sets of size $r$ is equivalent to picking the zero set of $g$  at random among all index sets of size $N-r$. The proof now makes use of Theorem \ref{theorem:rand-samp} to get a lower bound on $|\Omega(\mathcal{Z}(g))|$. For this we need to choose $d, \delta$ so that
\begin{equation}
\label{eq:ran-uncen-repeat}
N \log (1/\lambda) = N \log N/r > (1+\delta)d \log d.
\end{equation}
Fix any $\delta>0$ and let $d = N \log (N/r)/(1+\delta)\log N$. We check that $d, \delta$ satisfy $\eqref{eq:ran-uncen-repeat}$:
\begin{align*}
(1+\delta)d\log d &= \frac{N \log (N/r)}{\log N} \log \left(\frac{N \log (N/r)}{(1+\delta)\log N} \right)\\
&<\frac{N \log (N/r)}{\log N} \log N = N \log N/r,  
\end{align*}
Then  from Theorem  \ref{theorem:rand-samp}, 
\[
|\Omega(\mathcal{Z}(g))| \geq N\log (N/r) / (1+\delta)\log N 
\] 
with probability $1-d^{-\delta}$. From the uncertainty principle Theorem \ref{theorem:uncertainty}, we now have
\[
\begin{aligned}
|\text{supp}(\mathcal{F}g)| &\geq 1 + |\Omega(\mathcal{Z}(g))|\\
& \geq 1 + N\log (N/r) / (1+\delta)\log N 
\end{aligned}
\]
with probability  $1-d^{-\delta}$.

The final step in the proof uses a lower bound on $d = N \log (N/r)/(1+\delta)\log N$. We have set apart this technical result as Lemma \ref{lem: rnd-ucen-cvx}, below. This gives
\[
|\text{supp}(\mathcal{F}g)| \geq 1 + a_{N,\delta} - r \quad
\]
with probability $1-d^{-\delta}$.
Since $1-d^{-\delta} \geq 1- (a_{N,\delta} - r)^{-\delta}$, we can say
\[
|\text{supp}(\mathcal{F}g)| \geq 1 + a_{N,\delta} - r 
\]
with probability  $1-(a_{N,\delta} - r)^{-\delta}$.
The result follows since $ r = |\text{supp}(g)|$.
\end{IEEEproof}

\begin{lemma}
\label{lem: rnd-ucen-cvx}
Let 
\[
d = \frac{N\log(N/r)}{(1+\delta)\log N}
\]
and 
\[
a_{N, \delta} = \frac{N}{(1+\delta)\log N}\left(1+ \log (1+ \delta) + \log \log N\right),
\]
as  in Theorem \ref{thm:rnd-uncen}. Then $d \geq a_{N, \delta} - r$.
\end{lemma}

\begin{IEEEproof}
The convex function $\log (N/r)$ is bounded below by its tangent at any point $r_0>0$. Thus
\[
\log (N/r) \geq \log (N/r_0) + \left(-\frac{1}{r_0}(r - r_0)\right).
\]
For 
\[
r_0 = \frac{N}{(1+\delta)\log N}\,,
\]
 this reads
\[
\begin{aligned}
\log (N/r) &\geq \log \left((1+\delta)\log N\right)\\
& \hspace{.25in} + \left(-\frac{(1+\delta)\log N}{N}\left(r - \frac{N}{(1+\delta)\log N}\right)\right).
\end{aligned}
\]
Multiplying by $N/(1+\delta)\log N$, we have
\begin{align*}
d &= \frac{N\log(N/r)}{(1+\delta)\log N} \\
&\geq \frac{N\log \left((1+\delta)\log N\right)}{(1+\delta)\log N} - \left(r - \frac{N}{(1+\delta)\log N}\right) \\
&= \frac{N}{(1+\delta)\log N}\left(\log(1+\delta) + 1 + \log \log N \right) - r \\
&= a_{N, \delta} - r.
\end{align*}
\end{IEEEproof}

\begin{remark}
The robust uncertainty principle of Candes, Romberg and Tao in \cite{candes:robust} is as follows: for $M >0$ there exists a constant $C_M$ such that
\[
|\text{supp}(g)| + |\text{supp}(\mathcal{F}g)| \geq C_M N(\log N)^{-1/2},
\]
with probability $1- O(N^{-M})$.
This inequality is stronger than that of Theorem \ref{thm:rnd-uncen} by about $(\log N)^{-1/2}$. Also, Theorem \ref{thm:rnd-uncen} holds for $N=p^M$, whereas the inequality above holds for all $N$.

In our proof of Theorem \ref{theorem:rand-samp} we have only used the bound $|\Omega(\mathcal{Z}(g))| \geq p^k$ from Theorem \ref{thm:maximal-bounds}. By using the exact formula for $|\Omega(\mathcal{Z}(g))|$ in Theorem \ref{theorem:maximal-precise} (or by a better lower bound) it might be possible to tighten the uncertainty principle of Theorem \ref{thm:rnd-uncen} and remove the factor  $(\log N)^{-1/2}$.  
\end{remark}

\subsection{Sumsets and the Cauchy-Davenport Theorem}

Our final application is a  generalization of the Cauchy-Davenport theorem \cite{davenport:residues}, from additive number theory, on the size of sumsets. Again the inspiration comes from Tao's approach, \cite{tao:uncertainty},  to the original Cauchy-Davenport theorem via Chebotarev's theorem. 

\begin{theorem}
\label{thm:sumset-univ}
Let $\XX, \YY \subseteq [0:N-1]$. If either $\XX$ or $\YY$ is a universal sampling set, then 
\begin{equation}
\label{eq:sumset-univ}
|\XX+\YY| \geq |\XX| + |\YY| -1,
\end{equation}
when $|\XX| + |\YY| -1 \leq N$. 

Here $\XX+\YY$ is the sumset defined as
\[
\XX + \YY = \{x + y: x \in \XX, y \in \YY \},
\]
where the addition is modulo $N$. 
\end{theorem}

We are not assuming that $N$ is a prime power, while the classical theorem has $N=p$ and there are no assumptions on $\XX$ or $\YY$. 
That form of the result follows from Theorem \ref{thm:sumset-univ}, since all index sets in $[0:N-1]$ are universal when $N$ is prime. 

As a corollary we get a statement on the size of $|\XX+\YY|$ without making an assumption on $\XX$ or $\YY$.
\begin{corollary}
\label{cor:sumset-1}
Let $\XX, \YY \subseteq [0:N-1]$ be index sets. Then, 
\begin{equation}
\label{eq:sumset-gen}
|\XX+\YY| \geq \max\{ |\Omega (\XX)| + |\YY| -1, |\XX| + |\Omega (\YY)| -1 \}.
\end{equation}
\end{corollary}
\begin{IEEEproof} Since $\Omega(\XX) \subseteq \XX$, it follows that $\Omega(\XX) + \YY \subseteq \XX + \YY$. Now,
\begin{align*}
|\XX+ \YY| \geq |\Omega(\XX) + \YY| \geq |\Omega(\XX)| + |\YY| -1 \text{ from Theorem \ref{thm:sumset-univ}}.
\end{align*}
The inequality $|\XX + \YY| \geq |\XX| + |\Omega(\YY)| -1 $ follows similarly.
\end{IEEEproof}

\begin{IEEEproof}[Proof of Theorem \ref{thm:sumset-univ}]
First note that \eqref{eq:sumset-univ} follows trivially when either $X$ or $Y$ is a singleton. (More precisely, if, say,  $\XX$ is a singleton, then $\XX+\YY$ is just a translate of $\YY$, and so \eqref{eq:sumset-univ} holds with equality). For the rest of the proof, we assume that $|\XX|, |\YY| \geq 2$. Let $|\XX|=r$, $|\YY| =s$.

Assume without loss of generality that $\XX$ is universal. Let
\[
f_1 \in \BB^{\XX} \text{ be such that }f_1(\left[1:r\right]) = (\underbrace{0, 0, \ldots, 0}_{r-1 \text{ times }}, 1). 
\]
Such an $f_1$ exists because the set $[1:r]$, as an index set of $r$ consecutive integers, is a universal sampling set, so is in particular a sampling set for $\BB^\XX$. Similarly let
\[
f_2 \in \BB^\YY \text{ be such that }f_2(\left[r:r+s-1\right]) = (\underbrace{0, 0, \ldots, 0}_{s-1 \text{ times }}, 1),
\]
 again possible because $[r:r+s-1]$ is a set of $s$ consecutive integers, and hence a sampling set for $\BB^\YY$. Note that $f_1f_2 \in \BB^{\XX+\YY}$ and so $|\XX+\YY| \geq  \supp(\mathcal{F}(f_1f_2))$. Note also that the zero set  $\mathcal{Z}(f_1f_2)$ of $f_1f_2$ contains $[1:r+s-2]$, and hence, since the latter is a universal sampling set, $ |\Omega \left(\mathcal{Z}(f_1f_2)\right)| \geq r+s-2 = |\XX|+|\YY|-2$.
 
 Now we apply the uncertainty principle of Theorem \ref{theorem:uncertainty} to $f_1f_2$. We have, so long as $f_1f_2 \neq 0$,
\begin{align} \label{eq:sumset-0}
|\XX+\YY| &\geq \supp\left(\mathcal{F}(f_1f_2)\right) \nonumber \\
&\geq 1 + |\Omega\left(\mathcal{Z}(f_1f_2)\right)| \nonumber \\
& \geq 1 + |\XX|+|\YY|-2 = |\XX| + |\YY| -1, 
\end{align}
So we have proved that $|\XX+\YY| \geq |\XX| + |\YY| -1 $ if we know that $ f_1 f_2 \neq 0$. 

For this, again from Theorem \ref{theorem:uncertainty} we have 
\[
|\mathcal{Z}(f_1)| \leq |\Phi(\supp(\mathcal{F}f_1))| - 1 \leq |\Phi(\XX)| - 1,
\]
since $f_1 \in \mathbb{B}^\XX$. But $\XX$ is universal, so $\Phi(\XX)=\XX$ and
\begin{equation}
\label{eq:sumset-1}
|\mathcal{Z}(f_1)| \leq |\XX| -1. 
\end{equation}
By definition of $f_1$, the set $[1:r-1]=[1:|\XX|-1]$ is already in $\mathcal{Z}(f_1)$. Together with \eqref{eq:sumset-1}, this implies that $f_1$ cannot have any more zeros. In particular, $f_1(r+s -1) \neq 0$. Since $f_2(r+s -1) = 1$, $f_1f_2$ cannot be identically zero and \eqref{eq:sumset-0} applies.   \end{IEEEproof}

An important generalization of the Cauchy-Davenport theorem to any finite abelian group, not necessarily of prime order, is due to Kneser, \cite{kneser:sumset}. 
\begin{theorem}[Kneser]
 Let $G$ be a finite abelian group. Let $A, B \subseteq G$ be non empty subsets of $G$. Let $H$ be the set of periods, defined by $H = \{h \in G : h + (A+B) = A+B \}$. (Thus $A+B$ is periodic if $H \ne \{0\}$.) 
 Then 
 \[
 |A+B| \geq |A| + |B| - |H|.
 \]
Hence unless $A + B$ is periodic, $|A+B| \geq |A| + |B| - 1$. 
\end{theorem}  

Though the form is similar, this result neither implies nor is implied by Theorem \ref{thm:sumset-univ}. We give two examples. Let $N=8$, $\XX= \{0,1\}$, $\YY = \{0,4\}$. Then $\XX$ is universal and $\XX + \YY = \{0, 1, 4, 5\}$ is periodic with period $4$. So Theorem \ref{thm:sumset-univ} applies, but Kneser's theorem does not. Next let $N=16$, $\XX = \{0,2\}, \YY = \{0,2,4\}$. Then $\XX+\YY = \{0,2,4,6,8,10\}$, which is not periodic, and neither $\XX$ nor $\YY$ is universal. So Kneser's theorem applies, but Theorem \ref{thm:sumset-univ} does not. We hope to understand this more thoroughly.

\appendices{\section{Condition Number Associated with the Universal Sampling Set $\II^*$} \label{appendix:condition-number}}

An index set of consecutive integers is the simplest universal sampling set, but there is a catch in using it.
Let $\II$ be a universal sampling set of size $d$, $f\in \mathbb{C}^N$, and $f_\II$ the $d$-vector obtained from $f$ by sampling at locations in $\II$. If $f$ is in some bandlimited space $\mathbb{B}^\JJ$, $|\JJ|=d$, then the interpolation formula \eqref{eq:interpolation-formula-2} reads
\[
f = \Fou E_\JJ (E_\II^T \Fou E_\JJ)^{-1} f_\II.
\]
The practical difficulty is the computation of the inverse of $E_\II^T \Fou E_\JJ$. Suppose we use $\II = \II^* =[0:d-1]$ as a universal sampling set. We give a lower bound on the condition number of $E_\II^T \Fou E_\JJ$ that can be quite large for some $\JJ$, even though the matrix $E_\II^T \Fou E_\JJ$ is invertible for all $\JJ$. 

For $\II = [0:d-1]$, note that
\begin{align*}
|\det\left(E_{\II}^T \mathcal{F} E_{\JJ} \right)| &= |\det(\zeta_N^{ij})_{i \in \II, j \in \JJ} |\\
&= \prod_{j_1, j_2 \in \JJ}|\zeta_N^{j_1} - \zeta_N^{j_2}| \\
&= \prod_{j_1, j_2 \in \JJ}\left|2\sin\frac{2\pi(j_1-j_2)}{N}\right|.
\end{align*}

If $\{\sigma_i\}$ are the singular values of $A = E_{\II}^T \mathcal{F} E_{\JJ}$, then 
\begin{equation}
\label{eq:minsgn}
\det(A) = \sigma_1\sigma_2\sigma_3\ldots \sigma_d \geq \sigma_{\min}^d.
\end{equation}
Also if $a_{rk} = \exp(-2\pi i rj_k/N)$ are the entries of $A$, then 
\begin{equation}
\label{eq:maxsgn}
d^2 = \sum_{r,k = 0}^{d-1}|a_{rk}|^2 = \text{tr}(A^*A) = \sum_{r=0}^{d-1} \sigma_r^2 \leq d\sigma_{\text{max}}^2,
\end{equation}
and so $\sigma_{\text{max}}^2 \geq d$. 

From \eqref{eq:minsgn} and \eqref{eq:maxsgn}, the condition number satisfies
\[
\frac{\sigma_{\max}}{\sigma_{\min}} \geq \sqrt{d}\left(\frac{1}{\prod_{j_1, j_2 \in \JJ}|2\sin\frac{2\pi(j_1-j_2)}{N}|}\right)^{1/2d}.
\]
A possible scenario may be when $d$ is very small and $N$ is very large. In this case, the condition number can be very large if the frequency slots $\JJ$ are clustered.

\section{Counting Bracelets} \label{appendix:bracelets}

Several of our results, Theorem \ref{theorem:universal} for example, depend only on the bracelet of an index set rather than on the index set itself. Thus it is useful to know how many bracelets there are and how to enumerate them. Counting  bracelets -- actually, multicolored bracelets -- is a standard application in combinatorics of the orbit stabilizer theorem, and the problem is treated in many places. Our situation is slightly different because we want a count that specifies the number of black beads  in a black-and-white bracelet, corresponding to the size of the index set that determines the locations of the black beads. Nevertheless, the orbit stabilizer theorem can still be applied, and we have the following results.

\begin{theorem} Let $\phi$ denote Euler's totient function.
When $N$ is odd, the number of black-and-white bracelets of length $N$ with exactly $d$ black beads is
\[
\begin{array}{ll}
\frac{1}{2}  { { {(N-1)}/{2}  } \choose {  d/2 } }       + \frac{1}{2N}    \sum_{k|N , k|d} \frac{\phi(k)}{N} { {N/k} \choose {d/k}} 
		& \quad \textrm{for even } d, \\ \\
\frac{1}{2}   { { {(N-1)}/{2}  } \choose {  ({d-1})/{2} } }      +  \frac{1}{2N}    \sum_{k|N , k|d} \frac{\phi(k)}{N} { {N/k} \choose {d/k}} 
		& \quad \textrm{for odd } d.
		\end{array} 
\]

When $N$ is even, the number of black-and-white bracelets of length $N$ with exactly $d$ black beads is
\[
\begin{array}{ll}
\frac{1}{2}  {{ N/2  } \choose {  d/2 } }   +  \frac{1}{2N}    \sum_{k|N , k|d} \frac{\phi(k)}{N} { {N/k} \choose {d/k}} 
		& \quad \textrm{for even } d, \\ \\
\frac{1}{2}  {{(N/2) - 1} \choose {{(d-1)}/{2} }}    +  \frac{1}{2N}    \sum_{k|N , k|d} \frac{\phi(k)}{N} { {N/k} \choose {d/k}} 
		& \quad \textrm{for odd } d.
		\end{array} 
\]	
\end{theorem}

We omit the proof; see \cite{Will:thesis}. An efficient algorithm for enumerating bracelets has been devised only recently by Sawada \cite{sawada:bracelets}. An algorithm for determining when two index sets are in the same necklace is due to J.P. Duval \cite{duval:necklace}. It can also be used for bracelets. See \cite{Will:thesis} for examples of both of these.

\section{Additional References} \label{section:additional-references}

Though our work has concerned discrete-time signals exclusively, there is also a notion of universal sampling sets for continuous-time signals. We will not give the definition; it is interesting and not clear what the relations between the two may be. Here we cite only a few sources, starting with the paper of Landau \cite{landau:density} that featured the renowned necessary density condition on sampling sets. More recently, many interesting results have been obtained by Olevskii and Ulanovskii \cite{olevskii-ulanovskii:universal}, \cite{olevskii-ulanovskii:universal-2} on universal sampling and stable reconstruction, by Matei and Meyer \cite{matei-meyer:variant-compressed}, who work with lattices and make contact with compressed sensing, and by Bass and Gr\"ochenig  \cite{Bass04randomsampling}, who consider random sampling. Of course, anyone writing on so fundamental a topic as sampling and interpolation will encounter an enormous literature, and most probably miss an equal or greater amount.  We apologize to the authors of works we have missed.

\section*{Acknowledgments}

There are many people to thank for their interest, insight, and encouragement over quite some time, in particular S. Boyd, M. Chudnovsky,  A. El Gamal, J.T. Gill, S. Gunturk, B. Hassibi, J. Sawada,  J. Smith, and M. Tygert.   We also thank the reviewers for their thorough and thoughtful comments.

\bibliographystyle{IEEEtran}




\begin{IEEEbiography}[{\includegraphics[width=1in,height=1.25in,clip,keepaspectratio]{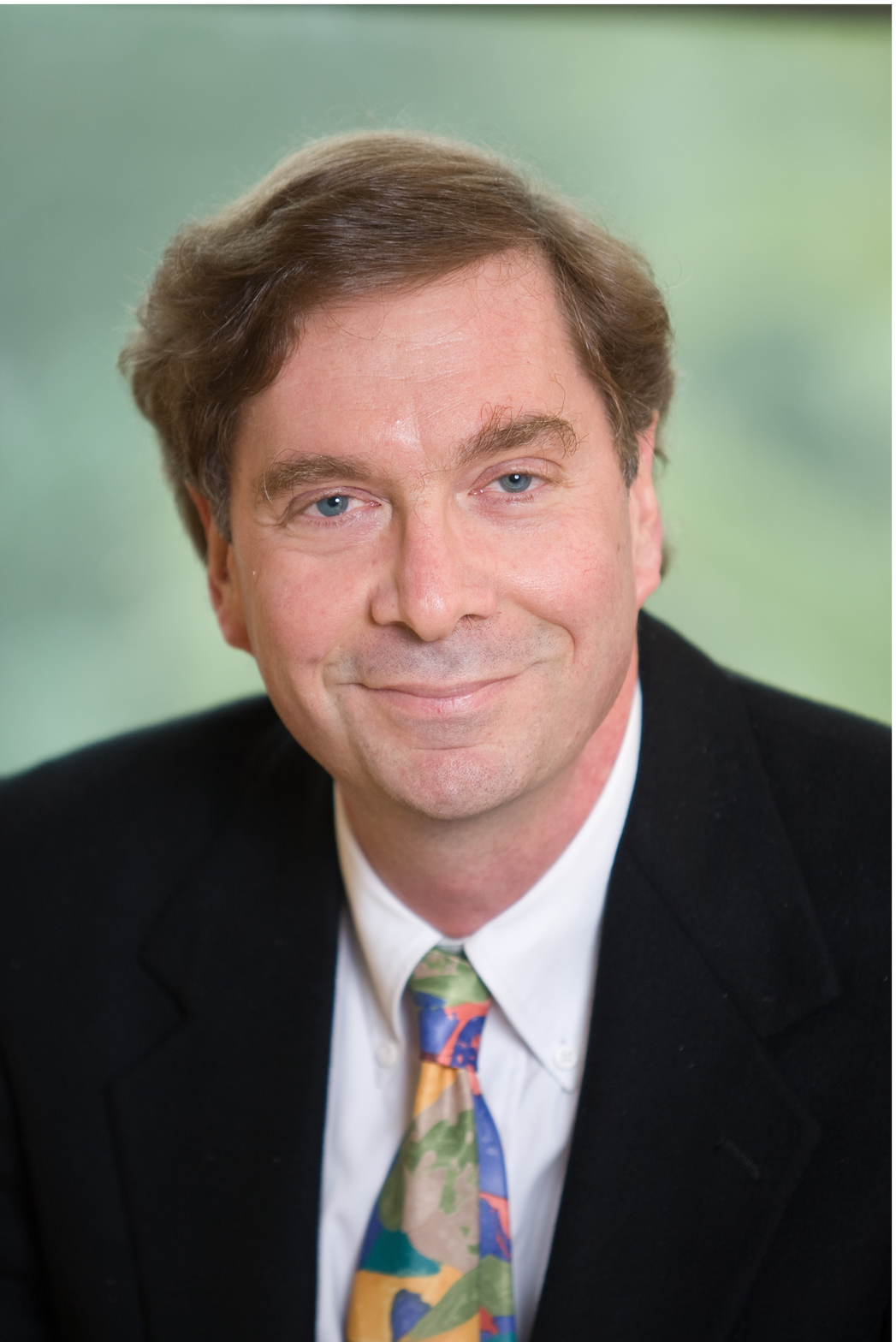}}]
{Brad Osgood} received his BS and MS at Carnegie-Mellon University and his Ph.D at the University of Michigan, all in mathematics. After a stint at Harvard he came to Stanford in 1985, first in the Mathematics Department and then in Electrical Engineering in the Information Systems Laboratory, where he is a Professor. Along with signal processing, his research in mathematics is in geometric function theory and differential geometry. Though becoming more digital, he plays trombone, the ultimate analog device.
\end{IEEEbiography}

\begin{IEEEbiography}[{\includegraphics[width=1in,height=1.25in,clip,keepaspectratio]{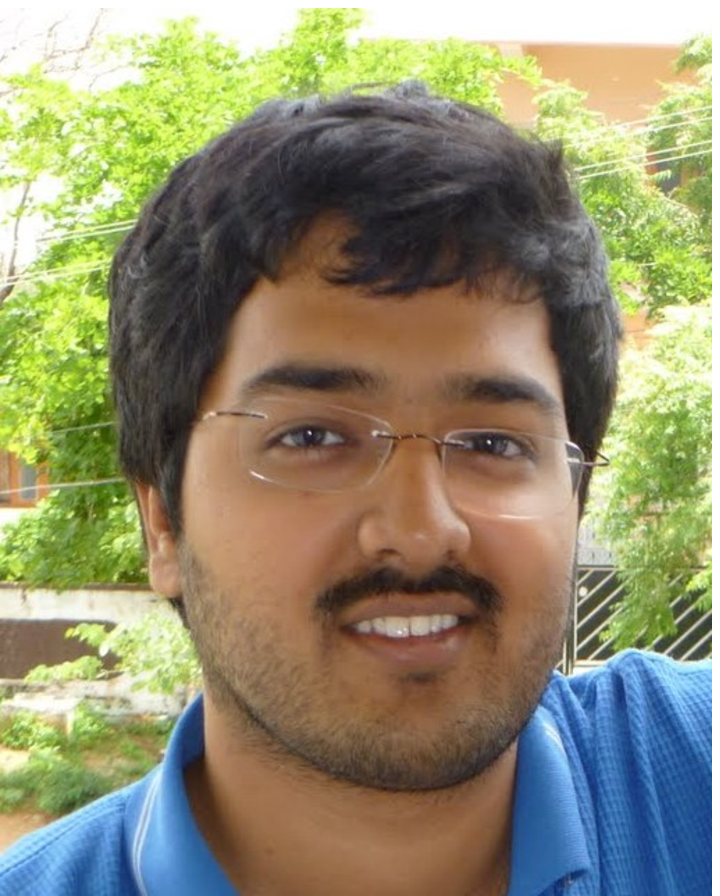}}]
{Aditya Siripuram} received his B.Tech and M.Tech degrees in Electrical
Engineering from Indian Institute of Technology, Bombay in 2009. He is
currently a PhD student in the Department of Electrical Engineering at
Stanford University, and a recipient of the Stanford Graduate
Fellowship. His interests include signal processing, coding
theory and recreational mathematics.

\end{IEEEbiography}

\begin{IEEEbiography}[{\includegraphics[width=1in,height=1.25in,clip,keepaspectratio]{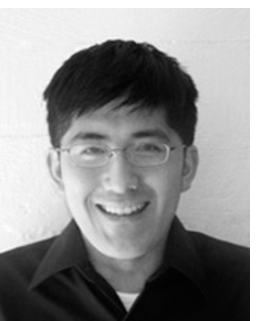}}]{William Wu}
received his B.Sc. in electrical engineering and computer science from
the University of California, Berkeley, and his M.Sc. in electrical
engineering, M.Sc. in mathematics, and Ph.D. in electrical engineering
from Stanford University. His dissertation focused on sampling and
reconstruction in finite dimensional signal spaces. Since 2010, he has
been a member of the technical staff at the Jet Propulsion Laboratory
(Pasadena, CA). His research interests include signal processing,
information theory, scientific computation, and recreational math; he
is the creator of {\tt wuriddles.com}, an archive of puzzles.
\end{IEEEbiography}

\end{document}